\documentclass[preprint,aps,12pt,notitlepage,nofootinbib,tightenlines]{revtex4-1}

\usepackage{epsfig}
\usepackage{slashed}
\usepackage{graphicx}
\usepackage{multirow}
\usepackage{color}
\usepackage{caption}
\usepackage{subcaption}
\usepackage{amsmath}
\usepackage{mathrsfs}
\usepackage{bm}
\usepackage{braket}
\usepackage{booktabs}
\usepackage{array}
\usepackage[mathscr]{euscript}
\usepackage[
			colorlinks=true,
			linkcolor=blue,
			urlcolor=red,
			citecolor=blue]{hyperref}
\usepackage{dcolumn}
\usepackage{slashed}       % slashed p
\usepackage{ulem}

\graphicspath{{figure/}}
\begin{document}
\title{Searching for axion-like particles at future electron-positron colliders}
\author{Hua-Ying Zhang}
\email{huayingheplnnu@163.com}
\author{Chong-Xing Yue}
\email{cxyue@lnnu.edu.cn}
\author{Yu-Chen Guo}
\email{ycguo@lnnu.edu.cn}
\author{Shuo Yang}
\email{shuoyanglnnu@163.com}
\affiliation{Department of Physics, Liaoning Normal University, Dalian 116029, China}
%\date{\today}
\begin{abstract}
We investigate the prospects for discovering  axion-like particles~(ALPs) via a light-by-light~(LBL) scattering at two colliders, the future circular collider~(FCC-ee) and circular electron-positron collider~(CEPC). The pro\texttt{mi}sing sensitivities to the effective ALP-photon coupling $g_{a\gamma\gamma}$  are obtained. Our numerical results show that the FCC-ee and CEPC might be more sensitive to the ALPs with mass 2 GeV $\sim$ 8 GeV than the LHC and CLIC.
\end{abstract}

\maketitle
%\newpage
%\tableofcontents
\newpage
\section{Introduction}
\label{sec:intro}
At present, the ``strong-CP'' is still one of the theoretical problems that the standard model (SM) doesn't explain. To solve it, many well-motivated extensions of the SM have been proposed. An attractive solution was proposed by Peccei-Quinn which is called Peccei-Quinn mechanism and predicts the existence of QCD axions~\cite{Peccei:1977hh,PhysRevLett.40.223,Wilczek:1977pj}. The axion-like particle (ALP) is a generalization of the QCD axion, which is predicted by new physics models with the breaking of global U$(1)$ symmetry~\cite{Svrcek:2006yi,Arvanitaki:2009fg,Cicoli:2012sz,Arias:2012az}.  The ALP interactions with the SM fermions and gauge bosons  arise through five-dimensional operators, and their masses can be treated independently of their couplings~\cite{Georgi:1986df}. While their couplings to the Higgs bosons are given by dimension-6 operators and higher~\cite{Bauer:2016zfj,Bauer:2017ris}. This property makes ALPs have much wider parameter space and hence generate rich phenomenology at low- and high-energy experiments.

The constraints on the couplings of ALPs to the SM fermions or bosons have been widely studied using various experimental data from particle physics, astroparticle physics and cosmology for example see Refs.~\cite{Fortin:2021cog,Agrawal:2021dbo,dEnterria:2020hqy} and their references.
Generally, the bounds depend on the ALP mass range considered. For example, the electroweak gauge boson Z decaying into two or three photons at $ e^{+}e^{-}$ colliders and hadron colliders can generate severe constrains on the ALP parameter space  \cite{Jaeckel:2015jla}.
 The mass of ALP and its coupling to two photons have already been constrained using LEP data \cite{Acciarri:1994gb, Anashkin:1999yse} via the process $e^{+} e^{-} \to \gamma^{*} \to a \gamma \to 3\gamma$ \cite{Mimasu:2014nea}.
The couplings of ALPs to the massive gauge bosons are constrained mainly due to loop induced processes~\cite{Gavela:2019wzg,Brivio:2017ije,Izaguirre:2016dfi}. The current limits provide valuable information for direct searching ALPs in running or future high- and low-energy experiments.

The properties of ALPs in high-energy collider experiments have been extensively studied~\cite{Bauer:2017ris, Wang:2021uyb, Bauer:2018uxu}. For example,  ALPs can be effectively detected at high energy colliders in a light-by-light (LBL) scattering~\cite{Baldenegro:2019whq,Baldenegro:2018hng,Inan:2020aal,Inan:2020kif}. Using the proton tagging technique, the LHC generally is more sensitive to the heavy ALP searched by LBL scattering than other processes. Especially, in the mass region 0.6 TeV $\sim$ 2 TeV, the lowest values of the coupling of ALP with a pair of photons in the range of 0.4 TeV $^{-1}\sim$ 0.06 TeV$^{-1}$~\cite{Baldenegro:2018hng}. It has been shown that the CLIC  with the initial unpolarized Compton backscattered (CB) photons and polarized CB photons can also be used to search ALPs~\cite{Inan:2020aal,Inan:2020kif}. The CLIC searches with the unpolarized photon-photon scattering obtain the bounds on ALP-photon coupling about between 1$\times$$10^{-2}$ TeV${^{-1}}$ and 3$\times$$10^{-4}$ TeV${^{-1}}$  for the ALP masses between 1 TeV $\sim$ 2.4 TeV~\cite{Inan:2020aal}, and the polarized bounds are about 1.5 times stronger than the unpolarized ones in the same mass interval~\cite{Inan:2020kif}.

It is well known that, compared to hadron colliders such as the LHC, lepton colliders have higher luminosity and more clean experimental environment. The future circular collider (FCC-ee)~\cite{Gomez-Ceballos:2013zzn,Abada:2019zxq,Abada:2019lih} and the circular electron-positron collider (CEPC)~\cite{CEPCStudyGroup:2018rmc,CEPCStudyGroup:2018ghi} are the unprecedented luminosity and energy frontier $e^{+}e^{-}$ colliders, which can not only study the SM observables at unprecedented accuracy, but will be very useful to discover the evidence of new physics beyond the SM. The proposed CEPC can operate at different center-of-mass (c.m.) energy ($\sqrt{s}$) of 91.2 GeV, 161 GeV and 240 GeV with the integrated luminosity ($\mathcal{L}$) in the range of 16 $\sim$ 2.6 ab$^{-1}$.
 The FCC-ee is designed to run at $\sqrt{s}=91\sim 365$ GeV with $\mathcal{L}$ in the range of 150 $\sim$ 1.0 ab$^{-1}$.

In this paper, we will investigate the possibility of detecting ALP via considering its single production induced by two-photon fusion at the FCC-ee and CEPC and focus on comparing the respective sensitivities to the ALP free parameters. This paper is structured as follows: In Sec.~\ref{theoretical}, we briefly review the couplings of ALPs with the SM particles and show the cross sections of the LBL scattering induced by ALPs at $e^+e^-$ colliders. In Secs.~\ref{(A)} and \ref{(B)}, we analyze the possibility of detecting ALPs at FCC-ee and CEPC by Monte Carlo simulation. Finaly, Sec.~\ref{conclusions} includes our conclusions and a simple discussion.

\section{Effective couplings of ALP and its induced LBL scattering } \label{theoretical}

As ALPs come from the breaking of a global symmetry at high energy scale, their interactions with ordinary particles can be suitably described via an effective Lagrangian and studied in effective field theory framework~\cite{Georgi:1986df,Bauer:2017ris,Brivio:2017ije}. At low-energies, the general dimension-5 effective Lagrangian describing the interactions of ALP  with ordinary fermions and the gauge bosons is given by
\begin{eqnarray}\label{1L}
	\label{lagrangian}
	\begin{aligned}
		\mathcal{L}_{\text {eff }}^{D \leq 5}&= \frac{1}{2}\left(\partial_{\mu} a\right)\left(\partial^{\mu} a\right)-\frac{M_{a}^{2}}{2} a^{2}+\frac{\partial^{\mu} a}{\Lambda} \sum_{F} \bar{\psi}_{F} C_{F} \gamma_{\mu} \psi_{F} \\
		&+g_{s}^{2} C_{G G} \frac{a}{\Lambda} G_{\mu \nu}^{A} \tilde{G}^{\mu \nu, A}+g^{2} C_{W W} \frac{a}{\Lambda} W_{\mu \nu}^{A} \tilde{W}^{\mu \nu, A}+g^{\prime 2} C_{B B} \frac{a}{\Lambda} B_{\mu \nu} \tilde{B}^{\mu \nu},
	\end{aligned}
\end{eqnarray}
where ${ X}_{\mu\nu}$ denotes the field strength tensor, ${ \tilde X}^{\mu\nu}=\frac{1}{2} \varepsilon^{\mu \nu \alpha \beta} X_{\alpha \beta}$ with $\varepsilon^{0123}=1$ and $X\in\{G, B, W\}$. The coupling constants $g_{s}$, $g$ and $g^{\prime}$ correspond the groups $SU(3)_{C}$, $SU(2)_L$ and $U(1)_Y$, respectively. ${\Lambda} $ is the characteristic scale of global symmetry breaking and $M_{a} $ is the ALP mass.

After electroweak symmetry breaking, Eq.~\ref{1L} generates the $a\gamma\gamma$, $a\gamma Z$ and $aZZ$ couplings, which are related our calculation
\begin{eqnarray}\label{2L}
	\mathcal{L}_{\mathrm{eff}}^{D \leq 5} =  -e^{2} C_{\gamma \gamma} \frac{a}{\Lambda} F_{\mu \nu} \tilde{F}^{\mu \nu}-\frac{2 e^{2}}{s_{w} c_{w}} C_{\gamma Z} \frac{a}{\Lambda} F_{\mu \nu} \tilde{Z}^{\mu \nu}-\frac{e^{2}}{s_{w}^{2} c_{w}^{2}} C_{Z Z} \frac{a}{\Lambda} Z_{\mu \nu} \tilde{Z}^{\mu \nu},
\end{eqnarray}
with the Wilson coefficients reading
\begin{eqnarray}
	C_{\gamma \gamma}=C_{W W}+C_{B B}, \quad C_{\gamma Z}=c_{w}^{2} C_{W W}-s_{w}^{2} C_{B B}, \quad C_{Z Z}=c_{w}^{4} C_{W W}+s_{w}^{4} C_{B B},
\end{eqnarray}
where $s_{w}=\sin \theta_{w}$, $c_{w} = \cos \theta_{w}$ with $\theta_{w}$ being the weak mixing angle, and  $F_{\mu\nu}$ and $Z_{\mu\nu}$ are the field strength tensors of photon and Z boson, respectively.
In the following, the coupling parameter $g_{a\gamma \gamma}$ is defined as ${g_{a\gamma \gamma}} = 4  e^{2}\frac{C_{\gamma \gamma}} \Lambda$.
The relationship between $g_{a\gamma\gamma}$ and the ALP-photon coupling $f$ used in Refs.~\cite{Baldenegro:2018hng,Inan:2020aal,Inan:2020kif} is $\frac{g_{a\gamma \gamma}} 4 =f^{-1}$.

\begin{figure}[htb]
	\includegraphics [scale=0.5] {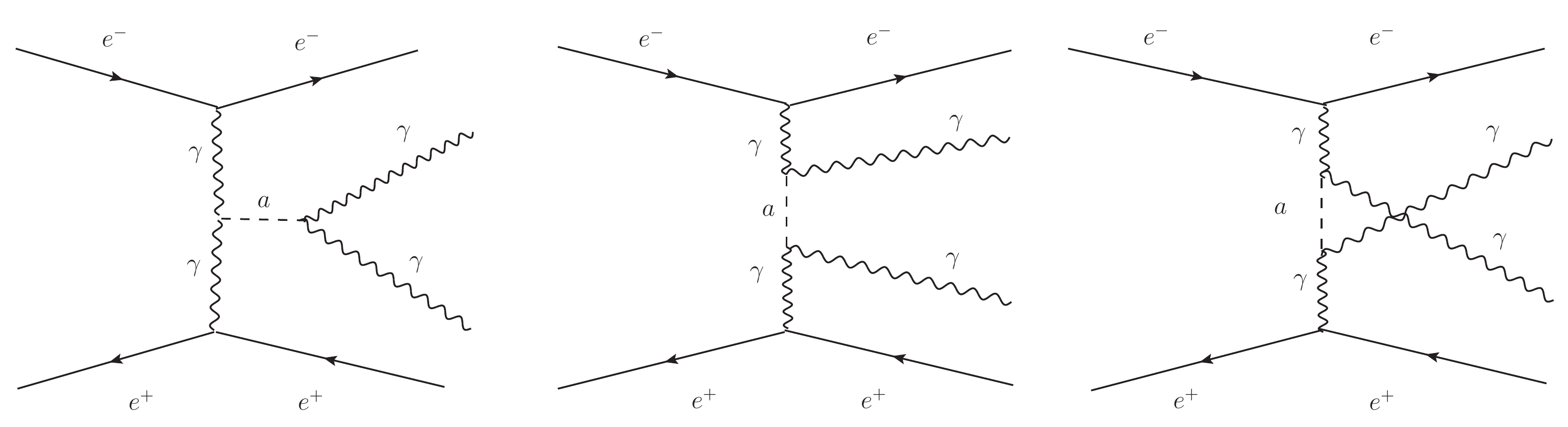}
	\caption{ The Feynman diagrams of the LBL scattering process induced by the $a\gamma\gamma$ coupling at $e^+e^-$ colliders.$~~~~~~~~~~~~~~~~~~~~~~~~~~~~~~~~~~~~~~~~~~~~~~~~~~~~~~~~~~~~~~~~~~~~~~~~~~~$ }
	\label{fig:Signal}
\end{figure}

In this paper, we consider the LBL scattering process induced by ALPs via $e^+e^-$ collision to search for ALPs.
The LBL scattering is characterized by the presence of a pair of photons and two forward electrons, which is shown in Fig.~\ref{fig:Signal}. Although this process at the CLIC has been studied in Refs.~\cite{Inan:2020aal,Inan:2020kif}, which shows the possibility of detecting ALPs at TeV. While ALPs have the potential to be found at the GeV scale. Thus, we focus our attention on the prospects for discovering ALPs at the FCC-ee and CEPC.

\begin{figure}[ht]
	\centering
	\begin{subfigure}{0.47\linewidth}
		\includegraphics[width=\linewidth]{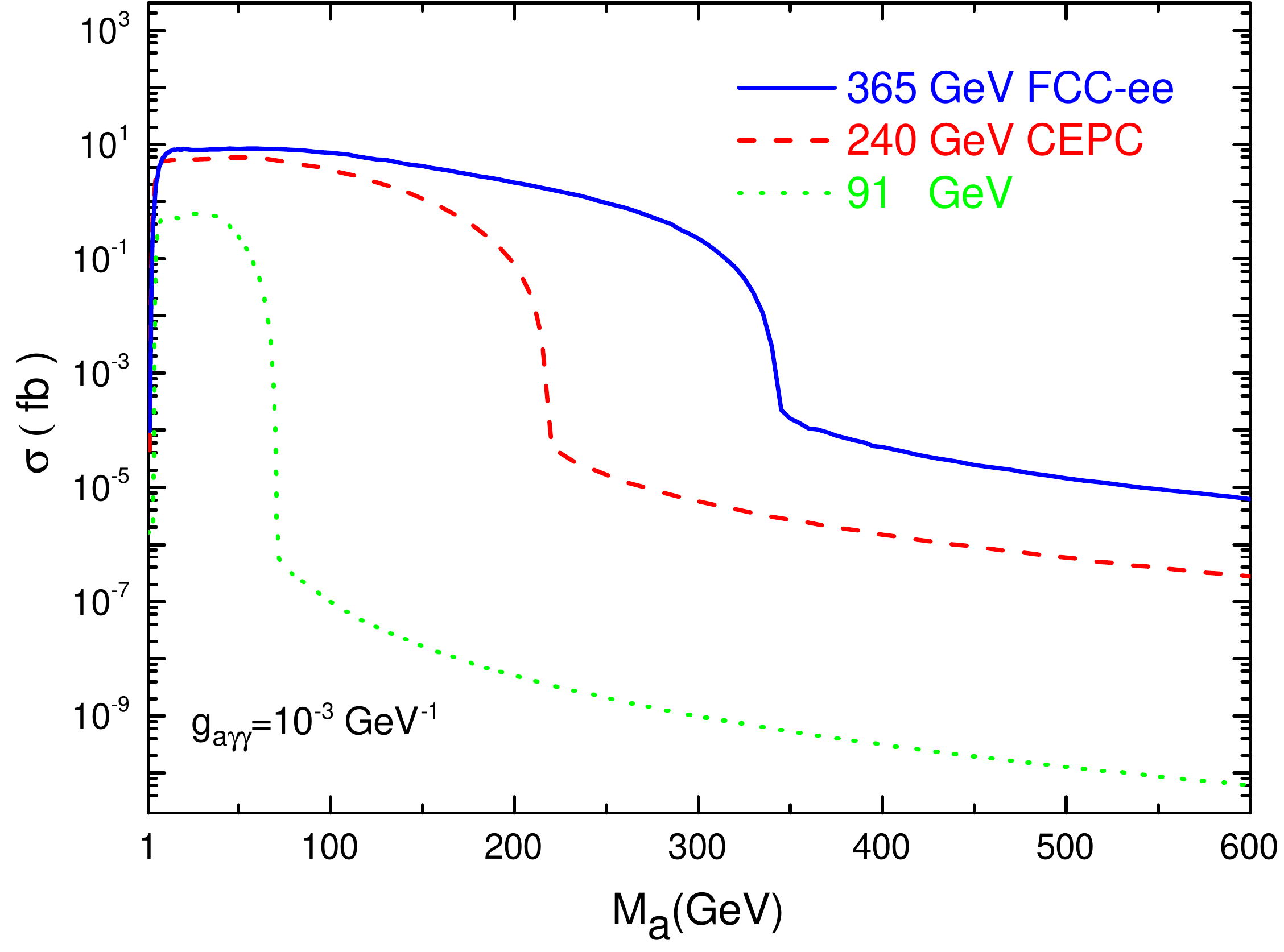}
		\caption{}\label{a}
	\end{subfigure}
	\begin{subfigure}{0.47\linewidth}
		\includegraphics[width=\linewidth]{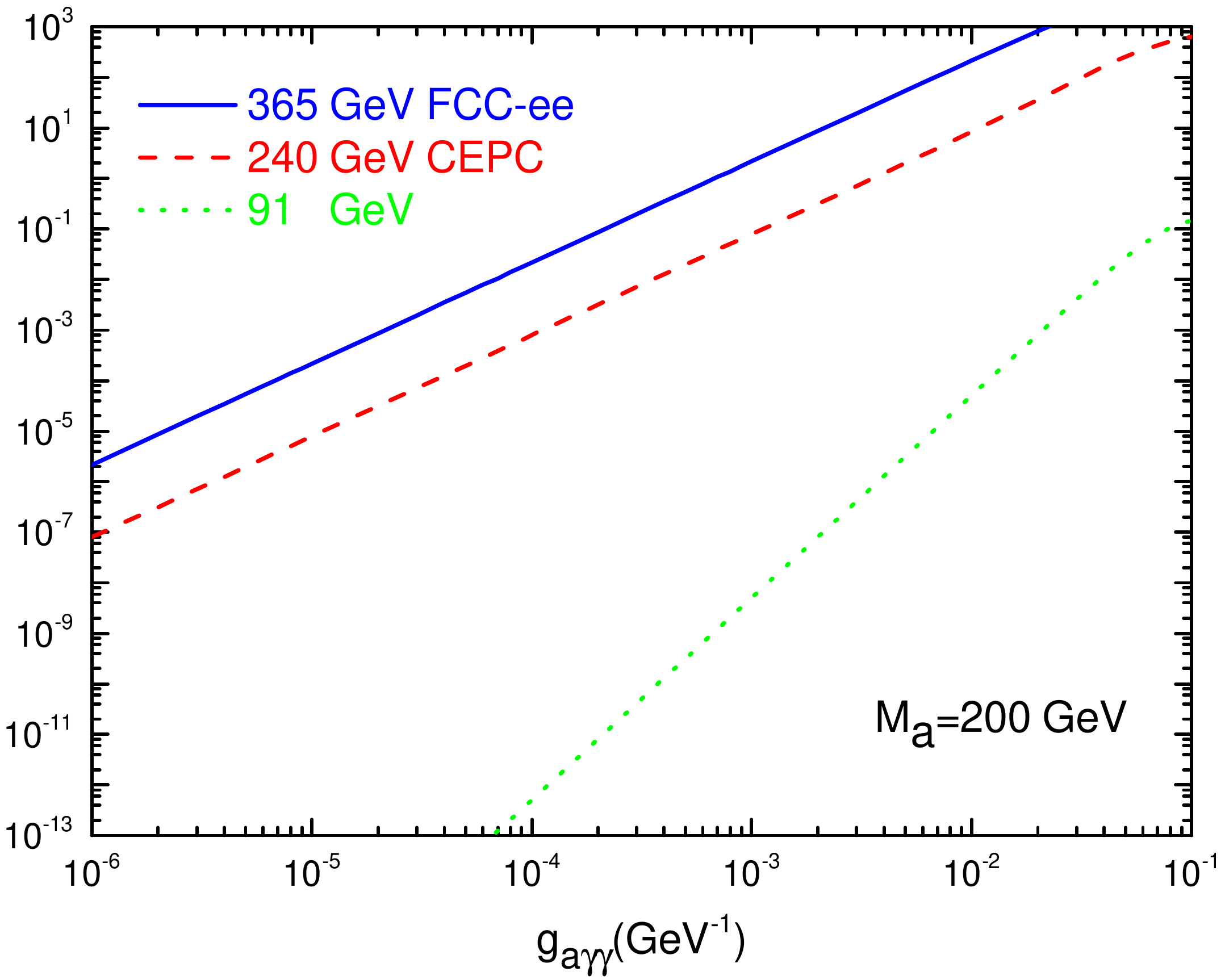}
		\caption{}\label{b}
	\end{subfigure}
	\caption{The production cross section $\sigma$ of the LBL scattering process induced by ALP as $ ~~ ~~~~~~$a function of the parameter $M_{a}$ or $g_{a\gamma\gamma}$ at 365 GeV FCC-ee (blue), 240 GeV CEPC (red) and 91 GeV (green).$ ~~ ~~~~~~~~~~~~~~~~~~~~~~~~~~~~~~~~~~~~~~~~~~~~~~~~~~~~~~~~~~~~~~~~~~~$
	}\label{scan}	
\end{figure}

We use FeynRules~\cite{Alloul:2013jea} to generate the Feynman rules corresponding to the effective Lagrangian. 
The cross sections of the process $e^+e^-\rightarrow \gamma \gamma e^+e^-$ induced by ALPs at the FCC-ee and CEPC are calculated by \textsf{Madgraph5-aMC@NLO}~\cite{madgraph} with the following set of basic cuts in Table \ref{basic cuts}, which are shown in Fig.~\ref{scan}.
 We choose the c.m. energies of FCC-ee as 365 GeV and 91 GeV~\cite{Gomez-Ceballos:2013zzn,Abada:2019zxq,Abada:2019lih}, and that of CEPC as 240 GeV and 91 GeV ~\cite{CEPCStudyGroup:2018rmc,CEPCStudyGroup:2018ghi}, respectively.
From Fig.~\ref{scan}, we can see that the values of the production cross sections decrease with the increase of the ALP mass $M_a$ and increase rapidly with the increase of the coupling parameter $g_{a\gamma\gamma}$.
%$$
\begin{table}[ht]\scriptsize  %\footnotesize {l}
	%{\color{red}
	\centering{
		\newcolumntype{C}[1]{>{\centering\let\newline\\\arraybackslash\hspace{0pt}}m{#1}}
		\begin{tabular}{C{2.4cm}|C{2.5cm}C{2.1cm}C{2.1cm}C{2cm}C{2cm}C{2cm} }
			%	\hline
			%	\multicolumn{7}{c}{ Basic cuts }\\
			\hline
			FCC-ee 365~GeV&$\Delta R(e, e)>0.4,$&$\Delta R(\gamma, e)>0.4,$&$\Delta R(\gamma, \gamma)>0.2,$&$\left|\eta_{e}\right|<2.5,$&$\left|\eta_{\gamma}\right|<2.4,$&$p_{T}^{i}>10~\mathrm{GeV}.$ \\
			
			FCC-ee 91~GeV$~$&$\Delta R(e, e)>0.4,$&$\Delta R(\gamma, e)>0.4,$&$\Delta R(\gamma, \gamma)>0.4,$&$\left|\eta_{e}\right|<2.4,$&$\left|\eta_{\gamma}\right|<2.4,$&$p_{T}^{i}>10~\mathrm{GeV}.$ \\
			
			$~$CEPC$~$ 240~GeV&$\Delta R(e, e)>0.4,$&$\Delta R(\gamma, e)>0.4,$&$\Delta R(\gamma, \gamma)>0.2,$&$\left|\eta_{e}\right|<2.4,$&$\left|\eta_{\gamma}\right|<2.4,$&$p_{T}^{i}>10~\mathrm{GeV}.$ \\
			
			$~$CEPC $~$91~GeV$~$&$\Delta R(e, e)>0.4,$&$\Delta R(\gamma, e)>0.4,$&$\Delta R(\gamma, \gamma)>0.4,$&$\left|\eta_{e}\right|<2.4,$&$\left|\eta_{\gamma}\right|<2.4,$&$p_{T}^{i}>10~\mathrm{GeV}.$ \\
			\hline   	
	\end{tabular}}
	\caption{The list of basic cuts. Here, $i = {\gamma}, e^{+}, e^{-}$, $\Delta R=\sqrt{(\Delta \phi)^{2}+(\Delta \eta)^{2}}$, $\Delta \phi$ and $\Delta \eta$ $~~~~~~~~~~~~~~$ are the azimuth difference and pseudo-rapidity difference of the photon/lepton pair or between a photon and a lepton, respectively.$~~~~~~~~~~~~~~~~~~~~~$  }\label{basic cuts}
	%}
\end{table}
 
\section{The possibility of detecting ALPs at the FCC-ee} \label{(A)}

\begin{figure}[htb]
	\includegraphics [scale=0.5] {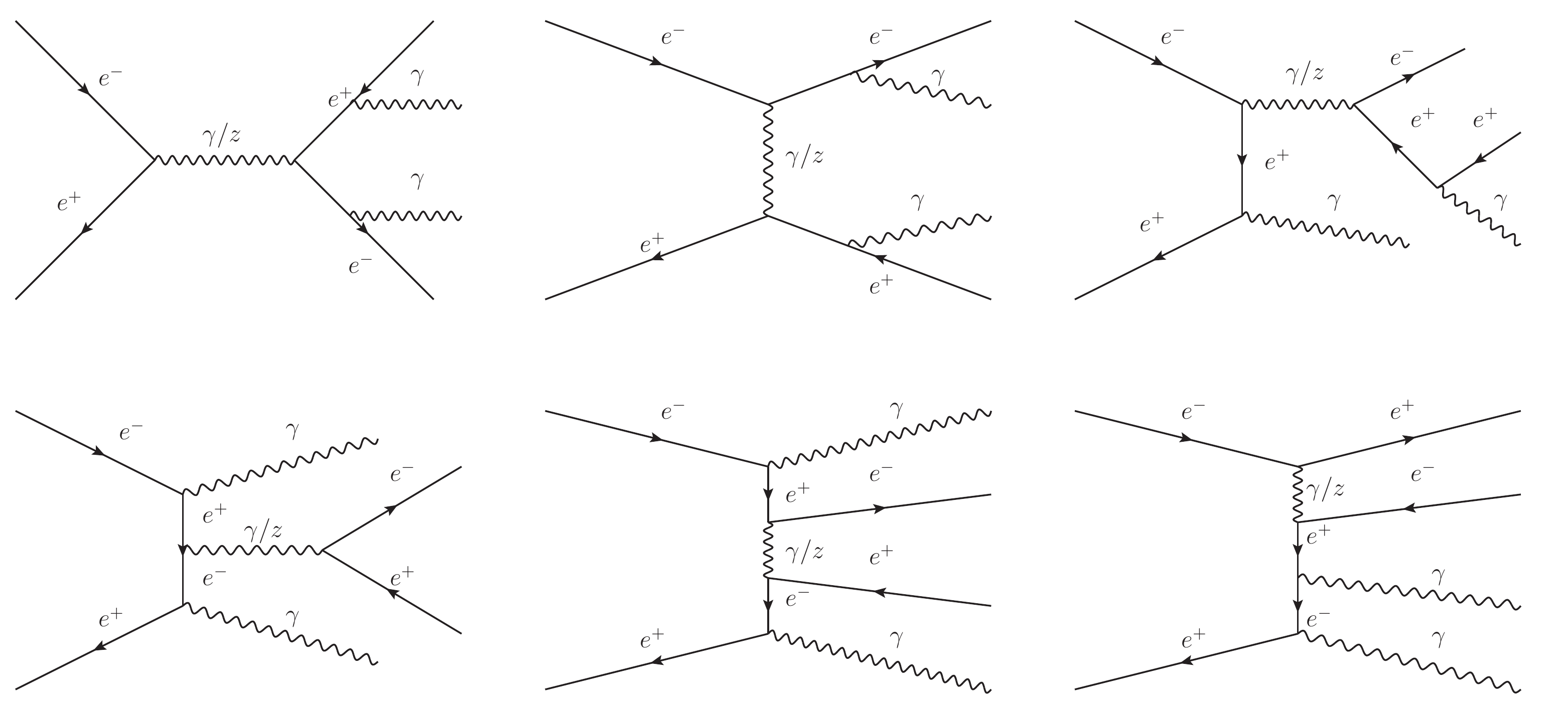}
	\caption{The typical diagrams for the background of the process $e^+ e^{-} \rightarrow\gamma \gamma e^+ e^{-}$.}
	\label{BGFeynman}
\end{figure}

We first study the feasibility of probing ALPs at the FCC-ee which will operate at $\sqrt{s}=365$ GeV with the integrated luminosity $\mathscr{L} = 1.0\ \mathrm{ab}^{-1}$ and $\sqrt{s}=91$ GeV with $\mathscr{L} = 150\ \mathrm{ab}^{-1}$ \cite{Abada:2019zxq}.
Compared with the LBL process, the contribution of the process $e^{+} e^{-} \to  a \gamma^{*}  \to \gamma \gamma e^{+} e^{-}$ induced by ALP is small. However, it can also contribute to this signal. So this process is considered as a supplement of ALP signal in the following analysis.
The dominant Feynman diagrams of the SM background are shown in Fig.~\ref{BGFeynman}, which are mainly induced by electroweak interaction.
The background also might be contaminated by the $\gamma\gamma$ initial state and $ZZ$-induced processes~\cite{Inan:2020aal}.
However, these possible backgrounds can safely be ignored since their contributions are estimated less than 1\%~\cite{Aad:2019ock,Atwood:1999cy}.

 In our simulation, the signal and background are first required to pass the set of basic cuts in Table \ref{basic cuts}.
With the help of the \textsf{Delphes}~\cite{deFavereau:2013fsa} and \textsf{MadAnalysis5}~\cite{Conte:2012fm} packages, we make fast detector simulations and cut-based analysis for the signal and background events.
In the simulation, the final states of the signal and background contain positron, electron and two photons. For the signal, the final state (two photons) from ALP decay could be a powerful trigger, and the angular separation of two photons depends largely on the ALP mass.
As a result, we choose the pseudo-rapidity of electron and positron $\eta {(e^{\pm})}$, the angle between reconstructed ALP and beam axis $\theta(\gamma \gamma)$, the angular separation between electron-positron $\Delta\theta(e^{+} e^{-})$, and transverse momentum of reconstructed ALP $p_{T} ({\gamma\gamma})$ as observables in the lab frame.

\begin{figure}[h!]
	\centering
	\begin{subfigure}{0.19\linewidth}
		\includegraphics[width=\linewidth]{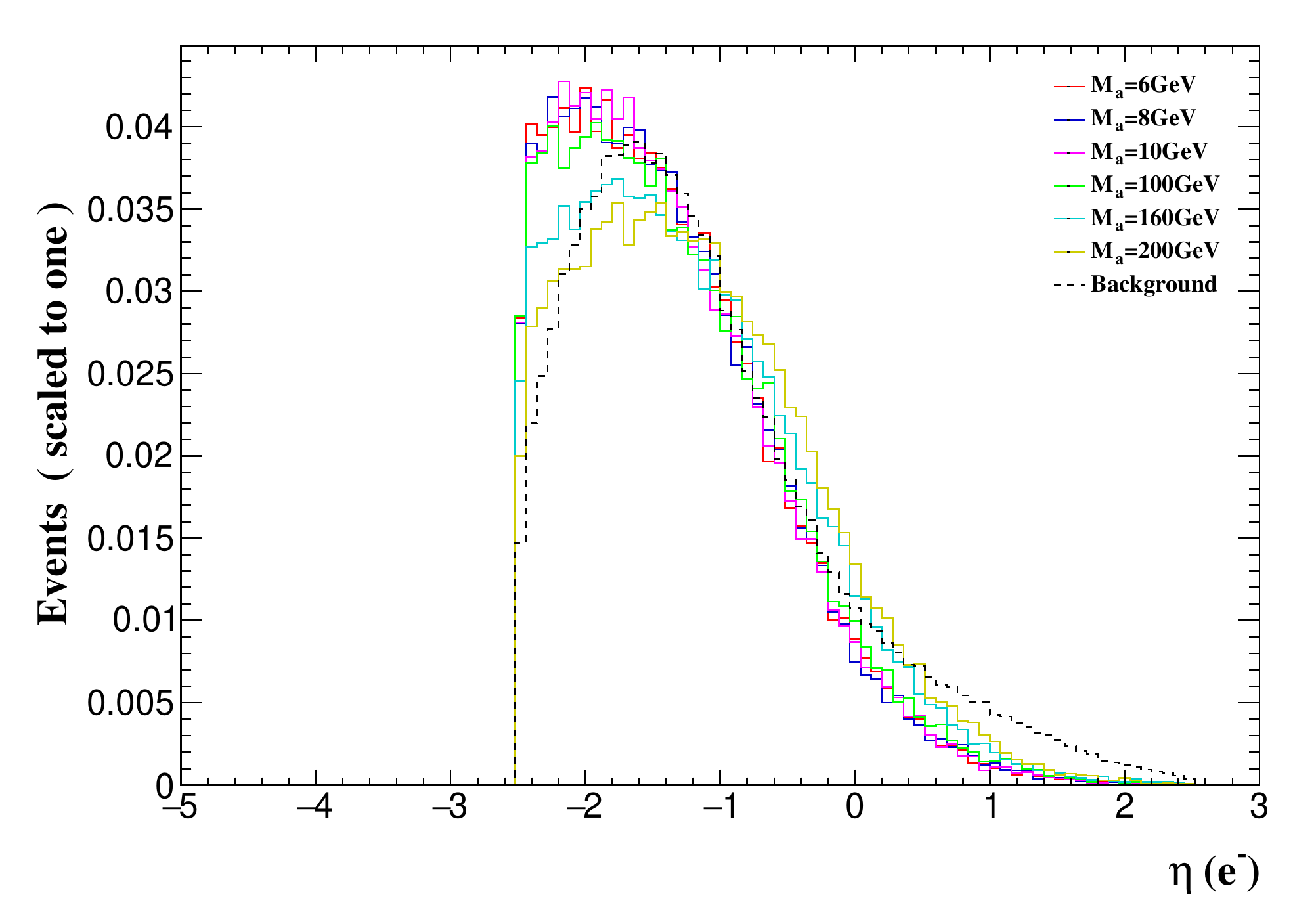}
		\caption{}\label{a}
	\end{subfigure}
\begin{subfigure}{0.19\linewidth}
	\includegraphics[width=\linewidth]{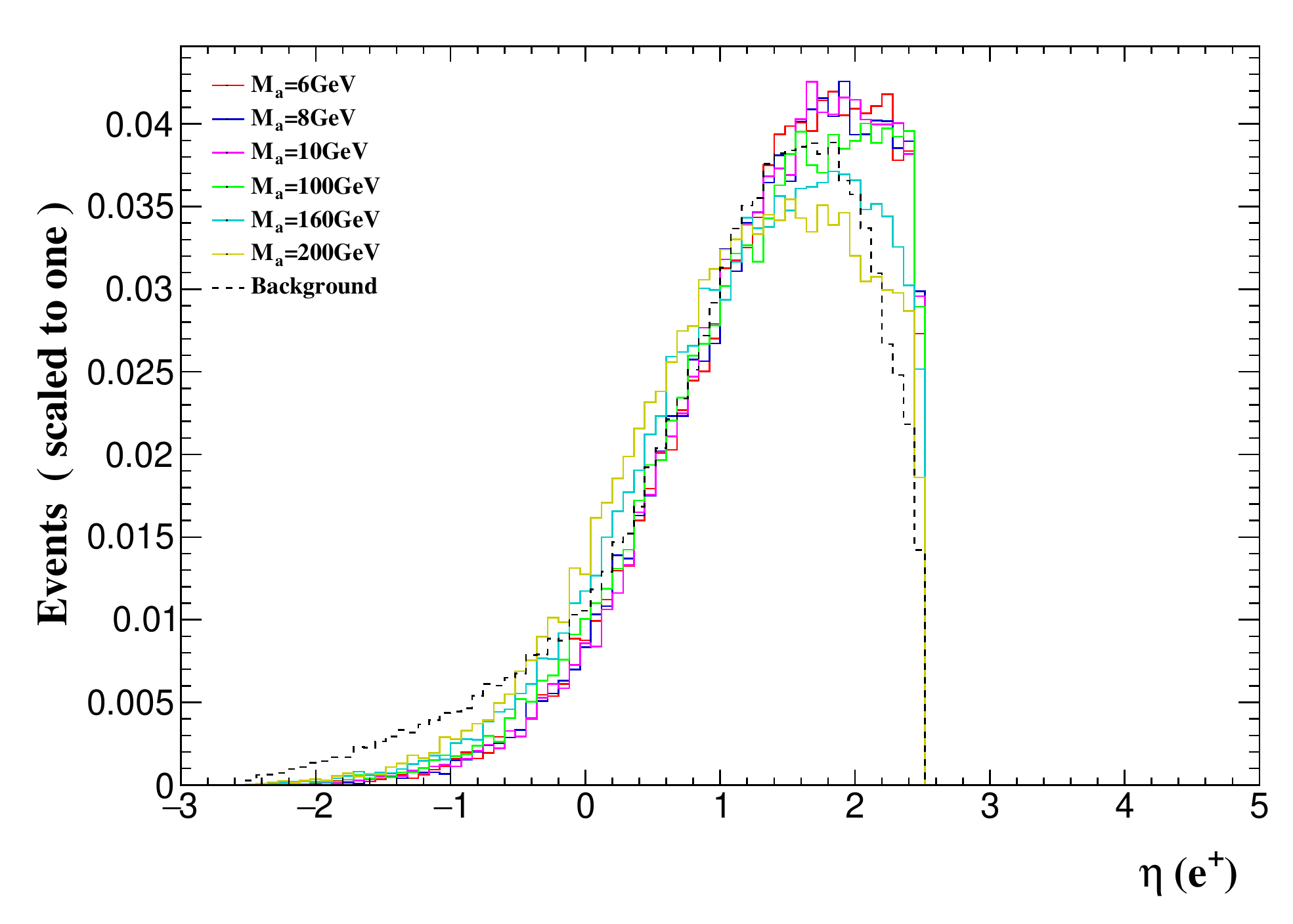}
	\caption{}\label{a}
\end{subfigure}
	\begin{subfigure}{0.19\linewidth}
		\includegraphics[width=\linewidth]{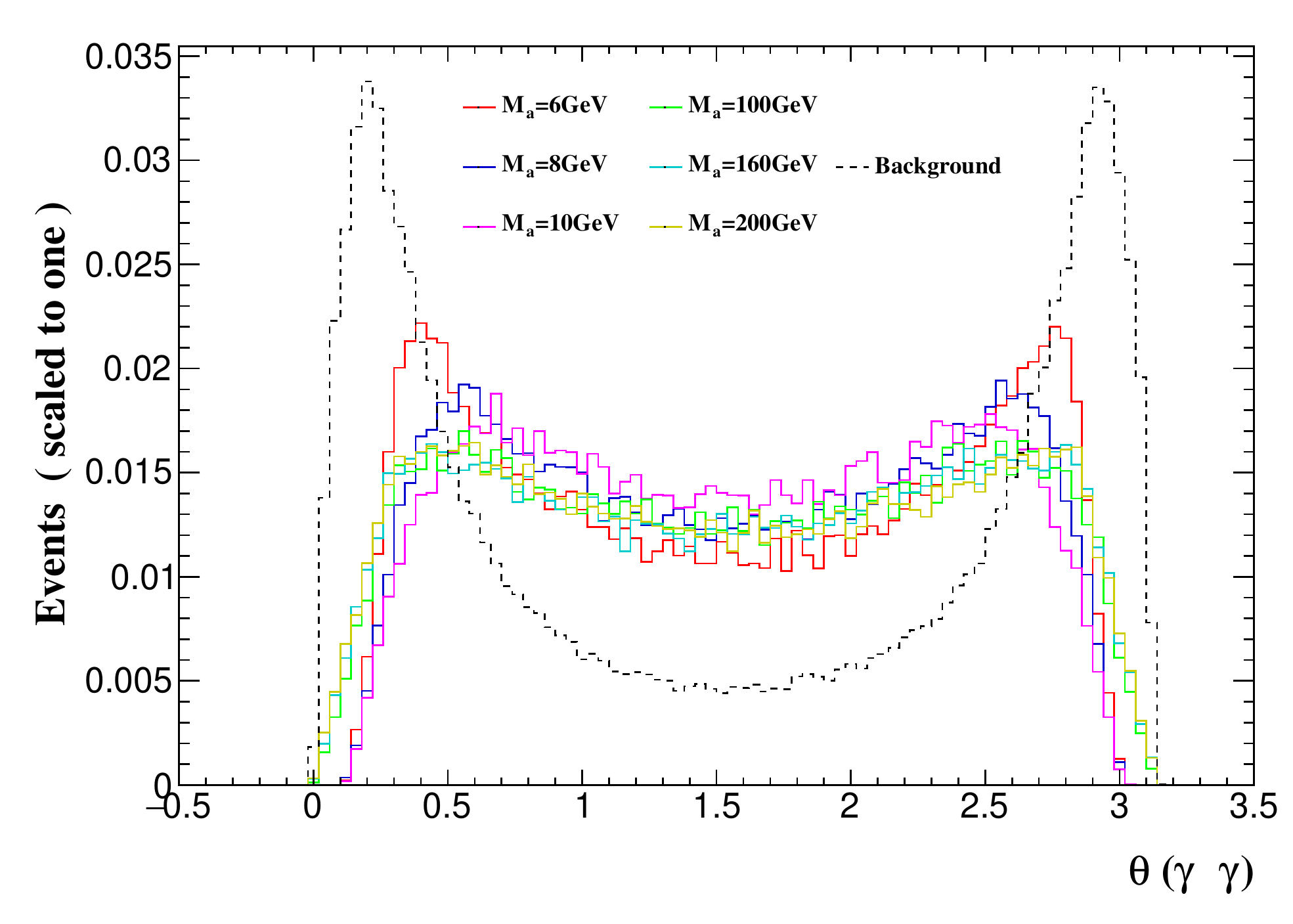}
		\caption{}\label{a}
	\end{subfigure}
	\begin{subfigure}{0.19\linewidth}
		\includegraphics[width=\linewidth]{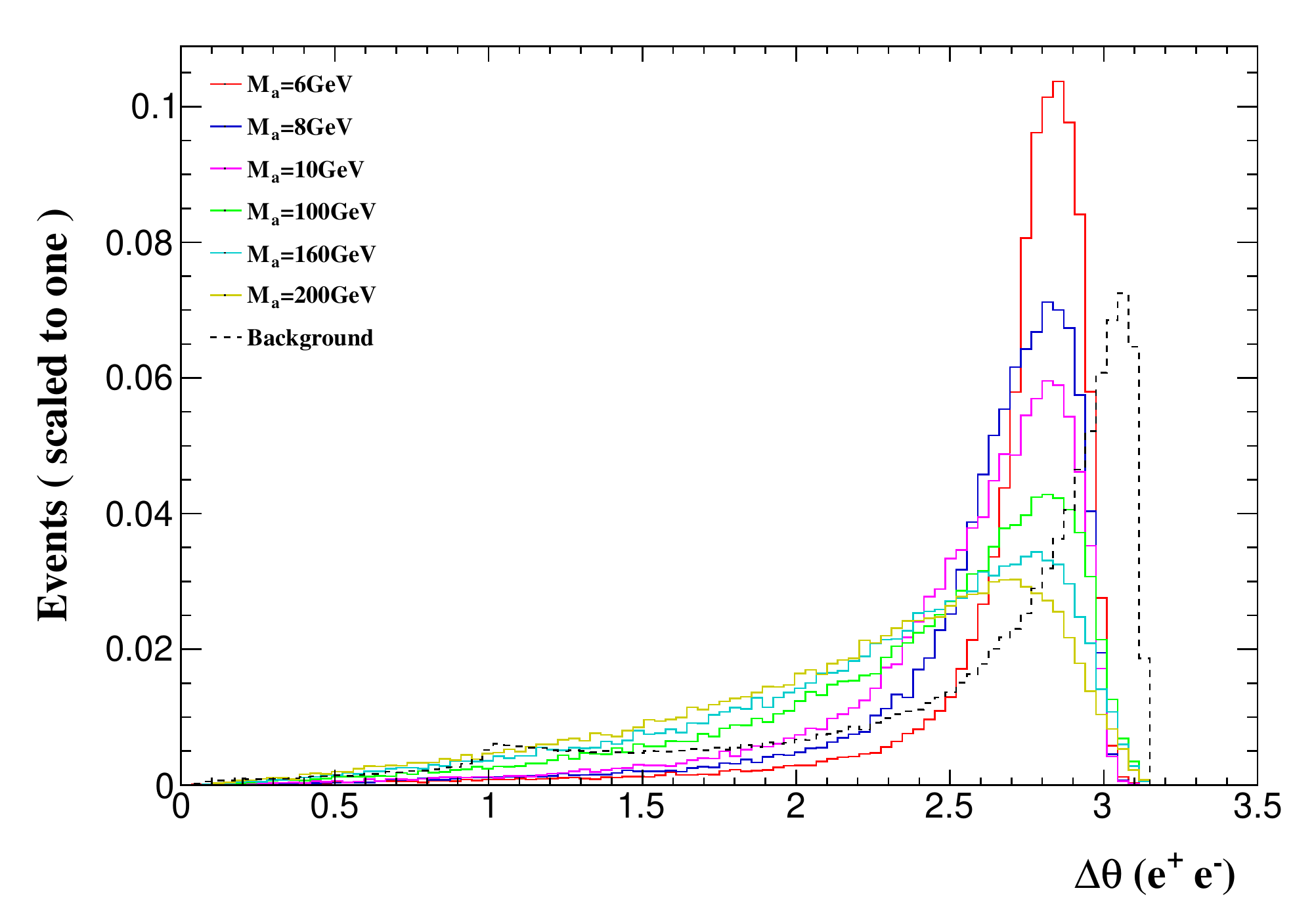}
		\caption{}\label{b}
	\end{subfigure}
	\begin{subfigure}{0.19\linewidth}
		\includegraphics[width=\linewidth]{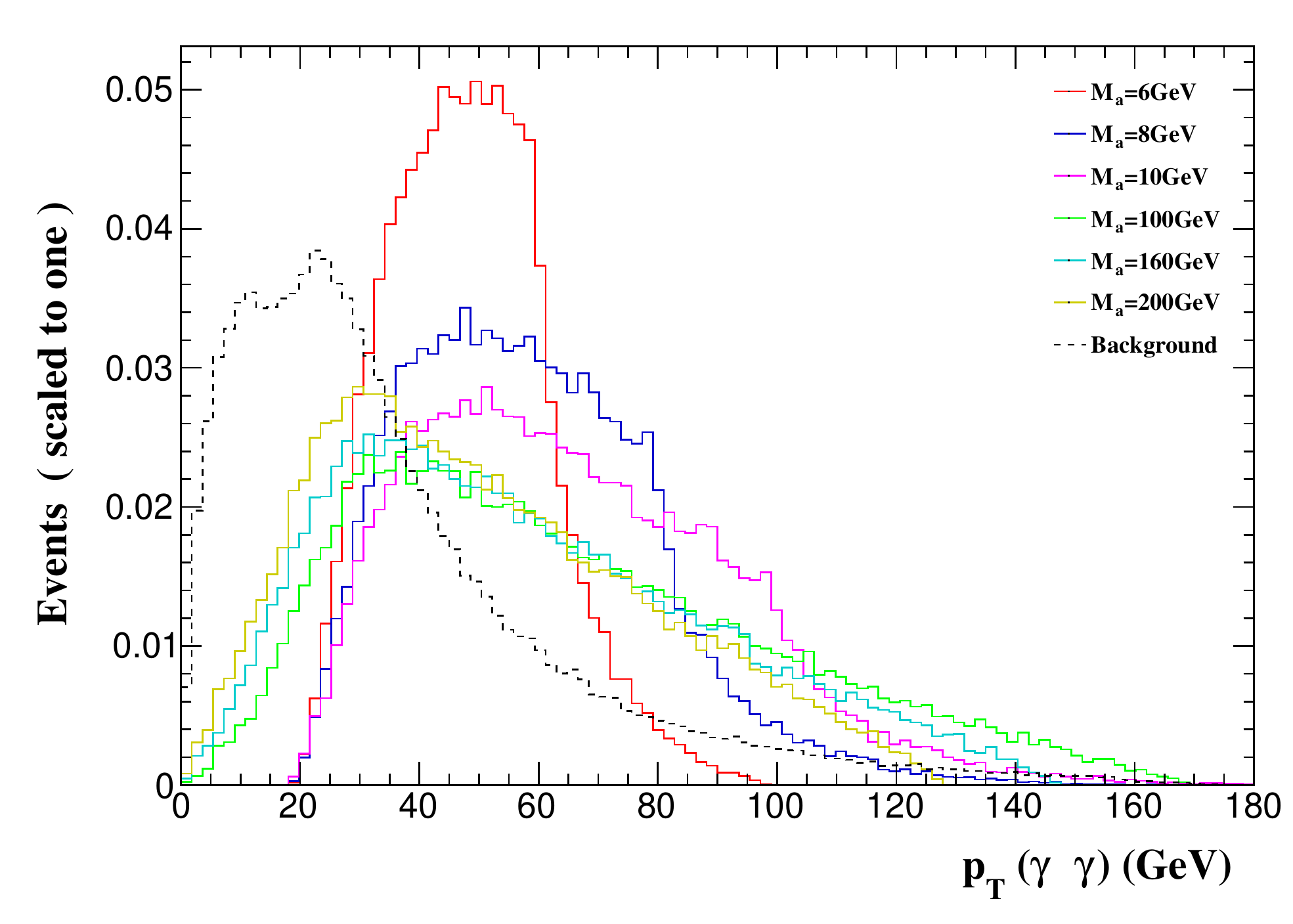}
		\caption{}\label{c}
	\end{subfigure}

	\begin{subfigure}{0.19\linewidth}
	\includegraphics[width=\linewidth]{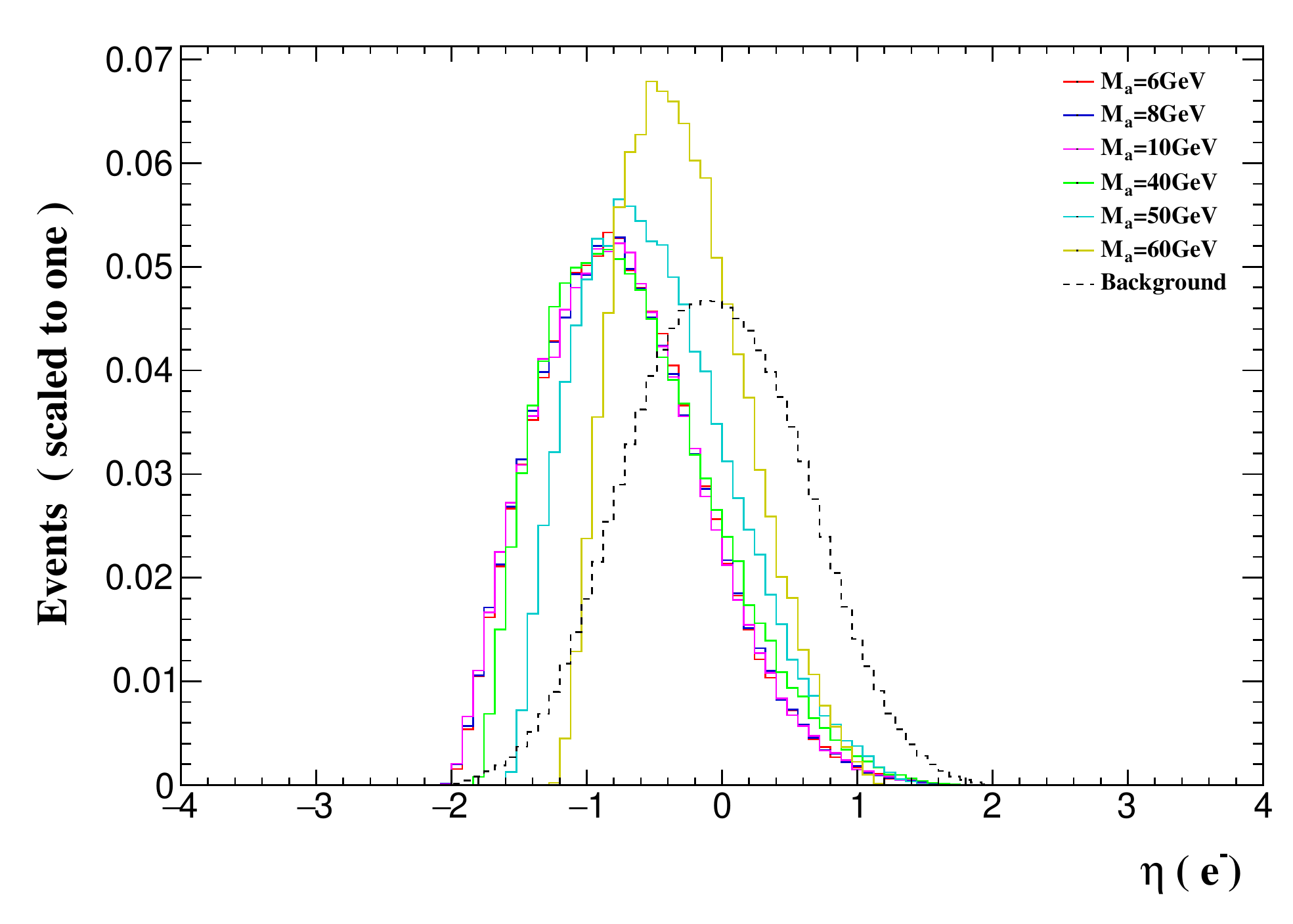}
	\caption{}\label{a.1}
	\end{subfigure}
	\begin{subfigure}{0.19\linewidth}
	\includegraphics[width=\linewidth]{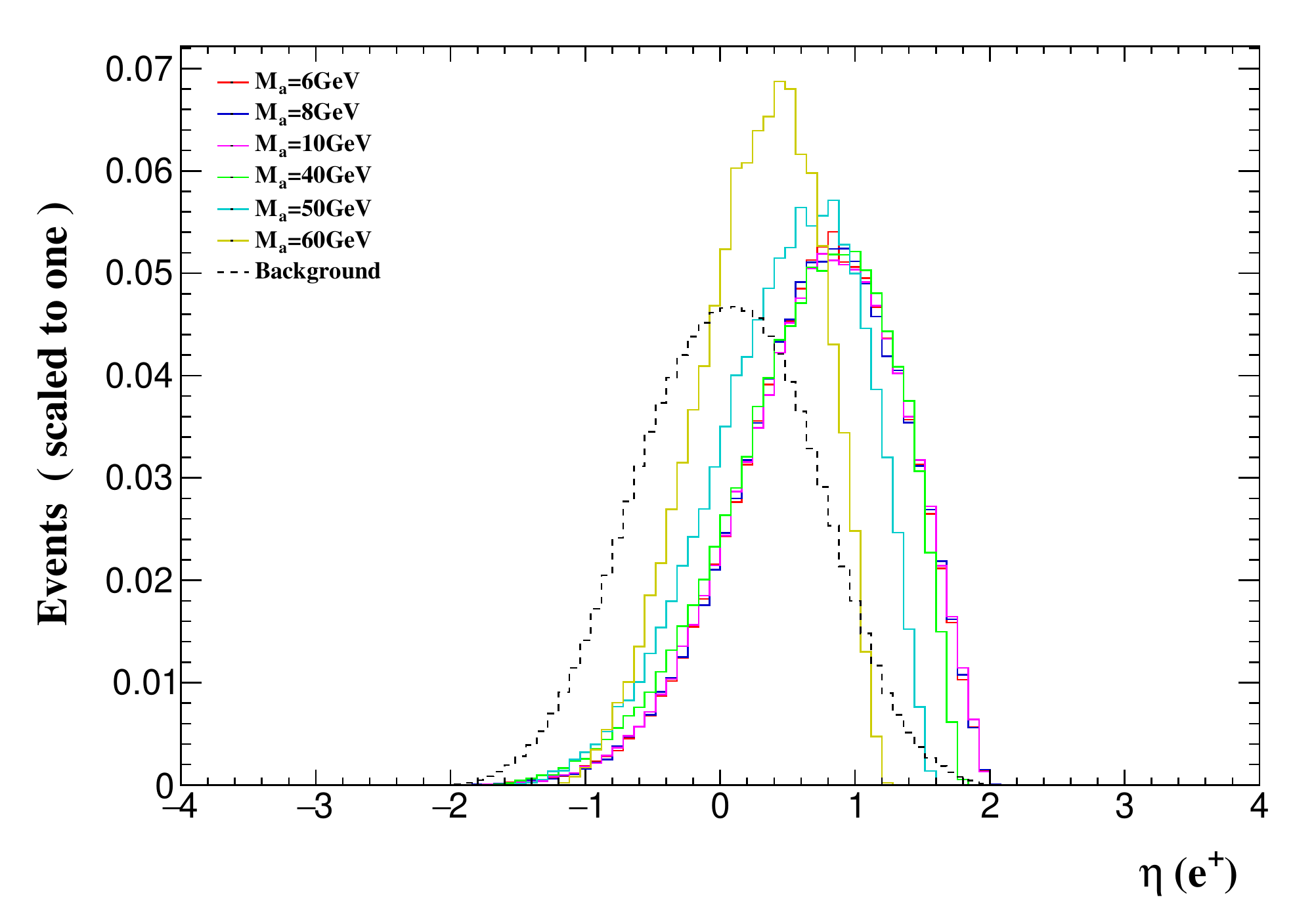}
	\caption{}\label{a.1}
	\end{subfigure}
	\begin{subfigure}{0.19\linewidth}
		\includegraphics[width=\linewidth]{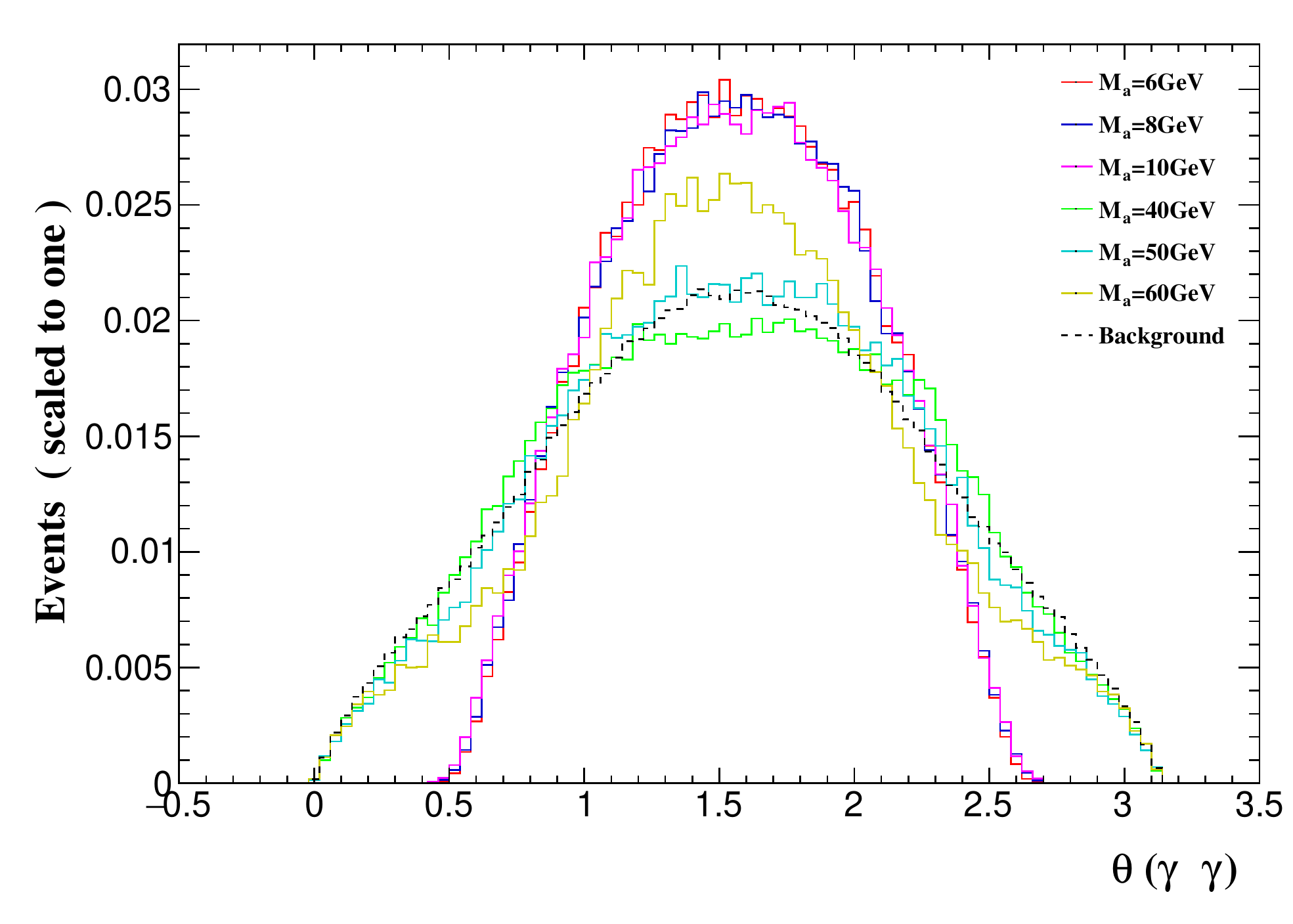}
		\caption{}\label{a.1}
	\end{subfigure}
	\begin{subfigure}{0.19\linewidth}
		\includegraphics[width=\linewidth]{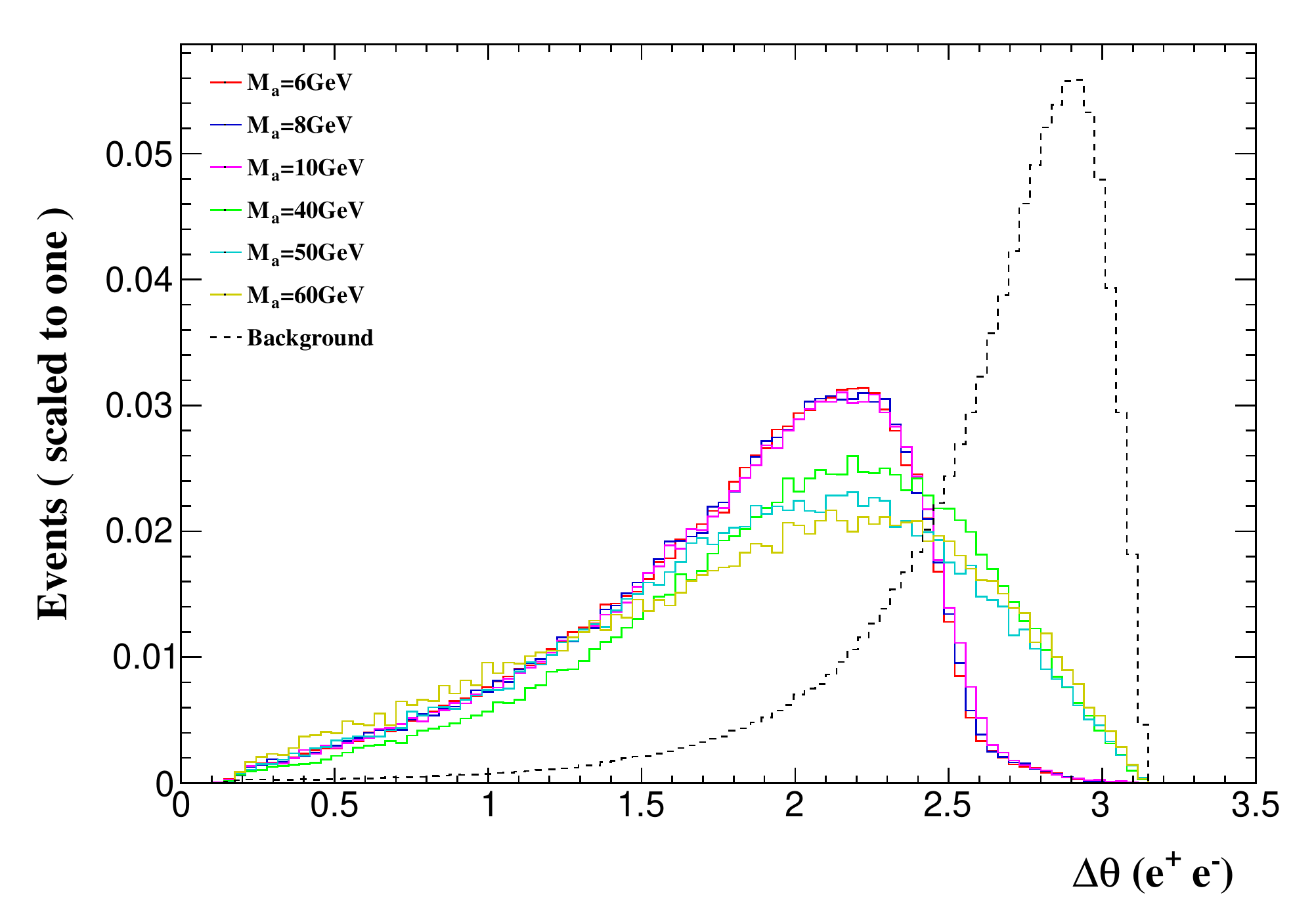}
		\caption{}\label{b.1}
	\end{subfigure}
	\begin{subfigure}{0.19\linewidth}
		\includegraphics[width=\linewidth]{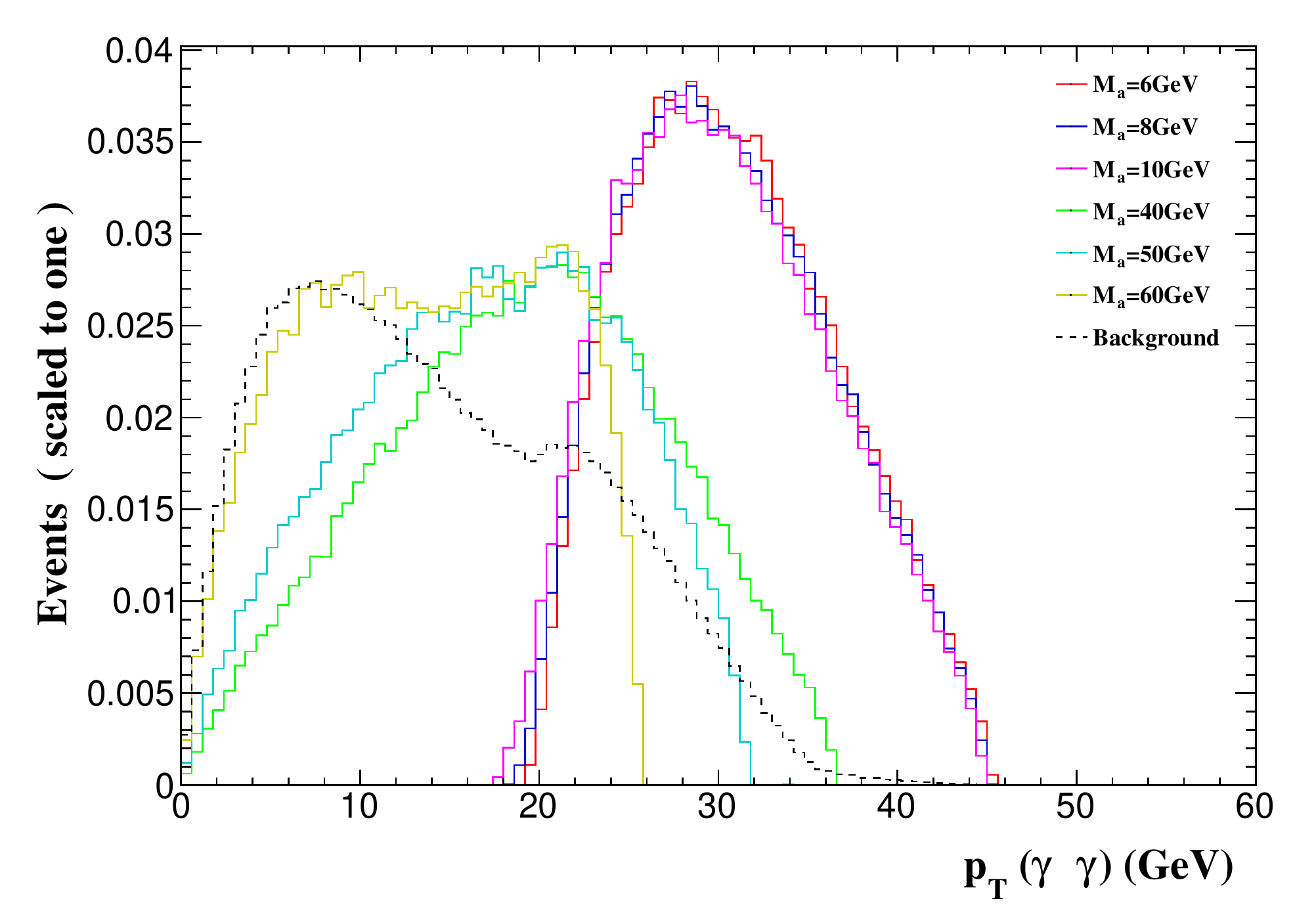}
		\caption{}\label{c.1}
	\end{subfigure}
	\caption{Normalized distributions of $\eta {(e^{\pm})}$, $\theta(\gamma \gamma),~\Delta\theta(e^{+} e^{-}),~p_{T} ({\gamma\gamma})$ for the signal of $~~~~~~~~~~~~~~~~$selected ALP masses and background at 365 GeV (a, b, c, d, e) and 91 GeV$~~~~$ $~~~~~$(f, g, h, i, j) FCC-ee with designed luminosities.$~~~~~~~~~~~~~~~~~~~~~~~~~~~~~~~~$}
	\label{fig3}
\end{figure}

In Fig.~\ref{fig3}, we show the normalized distributions of $\eta {(e^{\pm})}$, $\theta(\gamma \gamma)$, $\Delta\theta(e^{+} e^{-})$ and $p_{T} ({\gamma\gamma})$ of the signal and background at 365 GeV (91 GeV) FCC-ee for typical ALP masses of six benchmark points, $M_{a}$ = 6, 8, 10, 100, 160, 200 GeV (6, 8, 10, 40, 50, 60 GeV) with $g_{a\gamma\gamma}$= $10^{-3}$ GeV$^{-1}$. 
These normalized distributions are obtained after applying  the set of basic cuts in Table \ref{basic cuts}. 
 For the signal and background, the kinematic distributions are normalized to the number of expected events, which are the cross sections of the signal and background times integrated luminosities. 
From Figs.~\ref{fig3}~(a $\sim$ j), we can see that, in order to trigger the signal events, we need to impose some cuts on the relevant kinematic observables, which can help to suppress the background more effectively.
The expression of pseudo-rapidity is  $\eta(i) = -\ln  (\tan\dfrac{\theta_{i}}{2})$  ($i = e^{+}, e^{-}$). Here, $\theta_{i}$ is the angle between the emitted particle and the beam axis  in the lab frame.
For the LBL scattering process, most of positrons and electrons in the final state  still retain large energy. This will lead  $\theta_{e^{+}}$ is mainly concentrated in the range of $0^{\circ} \sim 90^{\circ}$ , while $\theta_{e^{-}}$ is mainly concentrated in the range of $90^{\circ} \sim 180^{\circ}$. Therefore the normalized distributions of $\eta(e^{+})$ and $\eta(e^{-})$ are asymmetrical and different in Fig. 4.
Comparing background with signal, we can find that the contribution of the $e^{+} e^{-} \to \gamma^{*} \to \gamma \gamma e^{+} e^{-}$ and weak processes to background is greater than that of the $e^{+} e^{-} \to  a \gamma^{*}  \to \gamma \gamma e^{+} e^{-}$ process to signal. Thus, for the normalized distributions of $\eta(e^{+})$ and $\eta(e^{-})$, there is a difference between the signal and the background. As can be seen from Figs. 4(f) and 4(g), this difference is more obvious at $\sqrt{s}=91$ GeV.
 Therefore, the restrictions can be imposed on $\eta({e^{+}})$ and $\eta({e^{-}})$ to increase the statistical significance~(SS). 
According to the above analysis,
we impose improved cuts on the signal and background in the following:
$$
\begin{array}{l}\scriptsize %\footnotesize
	\centering{
	\newcolumntype{C}[1]{>{\centering\let\newline\\\arraybackslash\hspace{0pt}}m{#1}}
	\begin{tabular}{C{8cm}C{3cm}C{3cm} }
			 Cuts     &$\sqrt{s}=$ 365~GeV & $\sqrt{s}=$ 91~GeV \\
		\multirow{2}{*}{\text { Cut-}1: Electron and positron pseudo-rapidity ~~~~~~~~~}&$0.6<\eta({e^{+}})<2.5 $&$-0.3<\eta({e^{+}})<0.9$ \\ 	
		                                                                                      &$-2.5<\eta({e^{-}})<-0.6$&$-0.9<\eta({e^{-}})<0.3$\\
	\text { Cut-}2: Angle between the ALP and the beam axis~~~~~ & 0.7~$\textless~\theta(\gamma \gamma)~\textless$~2.4 & ~~0.7~$\textless~ \theta(\gamma \gamma) ~\textless$~2.4 \\
	\text { Cut-}3:  Angular separation between electron-positron~~ &  $~~ \Delta\theta(e^{+} e^{-})~\textless$~2.9  &~~~  $\Delta\theta(e^{+} e^{-})~\textless$~2.4  \\
	\text { Cut-}4: Transverse momentum of reconstructed ALP~~~ &~~ $p_{T}({\gamma\gamma})~\textgreater~50 ~\mathrm{GeV}$ &~~~$p_{T}({\gamma\gamma}) ~\textgreater~20 ~\mathrm{GeV}$  \\	
	\end{tabular}}
\end{array}
$$

In Table~\ref{FCC_table}, we summarize the numerical results  of the signal and background after imposing above cuts at 365 (91) GeV FCC-ee.
For $\sqrt{s} = 365$ GeV, the signal benchmark points are taking as $M_{a} =$ 6, 8, 10, 50, 100, 200 GeV. For $\sqrt{s} = 91$ GeV and $g_{a\gamma\gamma}= 10^{-3}$ GeV$^{-1}$, the value of the signal cross section is smaller than 10 fb for $M_{a} >$ 60 GeV, therefore we take the signal benchmark points as $M_{a} =$ 6, 8, 10, 50 GeV.
From Table~\ref{FCC_table}, we can see that the background is suppressed very effectively, while the signal still has well efficiency after all cuts applied.
To estimate the SS, we use the following Poisson formula~\cite{Cowan:2010js}:
\begin{equation}\label{SS}
	SS = \sqrt{2 \mathscr{L} [(S + B)\ln(1+\frac{S}{B})-S]},
\end{equation}
where $S$ and $B$ respectively denote the effective cross sections of the signal and background after all cuts applied.

%%%%%%%%%%%%%%%%%%%%%%%%%%%%%%%%%%%%%%%%%%%%%%%%%%%%%%%%%%%%%%%%%%%%%%%%%%%%%%%%%%%%%%%%%%%%
%%%%%%%%%%%%%%%%%%%%%%%%%%%%%%%%%%%%%%%%%%%%%%%%%%%%%%%%%%%%%%%%%%%%%%%%%%%%%%%%%%%%%%%%%%%

\begin{table}[h!]\tiny
	\centering{
		\newcolumntype{C}[1]{>{\centering\let\newline\\\arraybackslash\hspace{0pt}}m{#1}}
		\begin{tabular}{|C{1.2 cm}|C{1.8cm}|C{1.8cm}|C{1.8cm}|C{1.8 cm}|C{1.8 cm}|C{1.8cm}|C{2.0cm}| }
			\hline
			\multicolumn{8}{|c|} { FCC-ee @ $\sqrt{s}=365$ (91) GeV  }\\
			\hline
			\multirow{2}{*}{Cuts}    & \multicolumn{6}{c|}{Signal (fb) }&\multicolumn{1}{c|}{Background (fb) }   \\%\multicolumn{3}{c|}{SS}
			\cline{2-8}
			&$M_a$ = 6~GeV  & $M_a$ = 8~GeV  & $M_a$= 10~GeV  & $M_a$= 50~GeV   & $M_a$= 100~GeV   & $M_a$= 200~GeV   & ${\gamma \gamma e^{+} e^{-}}$   \\
			\hline
			 Basic cuts &2.9092(0.2483)  &5.0074(0.4786) &6.5272(0.5001) &8.4206(0.2432)  &7.1235    &2.1737 &54.203(98.8188) \\
			Cut 1      &2.1634(0.0311)  &4.2978(0.1265) &5.3419(0.142) &4.5123( 0.0977)  &4.9093      & 1.2593 &29.233(41.0505)  \\
			Cut 2      &1.3962(0.0307)  &2.6956(0.1261) &3.6755(0.1416) &2.9963(0.0904)  & 3.1011     & 0.7942  & 8.3373(35.0206) \\
			Cut 3      &1.2374(0.0223)  &2.5417(0.1152) &3.5173(0.1304) &2.8482 (0.0717)  & 2.9926    & 0.768 & 4.8137(8.4019)  \\
			Cut 4      &0.9014(0.0222)  &2.2243(0.115) &3.1819(0.1303) &2.5198(0.05)     &  2.5458    &  0.453 & 2.6445(6.1842) \\
			\hline
	\end{tabular}}	
	\caption{After different cuts applied, the cross sections for the signal and background at $~~~~~~~~~~~~~~~$ 365 (91) GeV FCC-ee with $g_{a\gamma\gamma} = 10^{-3}$ GeV$^{-1}$ and the benchmark points $M_{a}$ = 6, 8, 10, 50, 100, 200 GeV (6, 8, 10, 50 GeV).$~~~ ~~~~~~~~~~~~~~~~~~~~~~~~~~~~$\label{FCC_table}}
\end{table}
%%%%%%%%%%%%%%%%%%%%%%%%%%%%%%%%%%%%%%%%%%%%%%%%%%%%%%%%%%%%%%%%%%%%%%%%%%%%%%%%%%%%%%%%%%%
%%%%%%%%%%%%%%%%%%%%%%%%%%%%%%%%%%%%%%%%%%%%%%%%%%%%%%%%%%%%%%%%%%%%%%%%%%%%%%%%%%%%%%%%%%%
\begin{figure}[!ht]
	\begin{center}
		\includegraphics [scale=0.4] {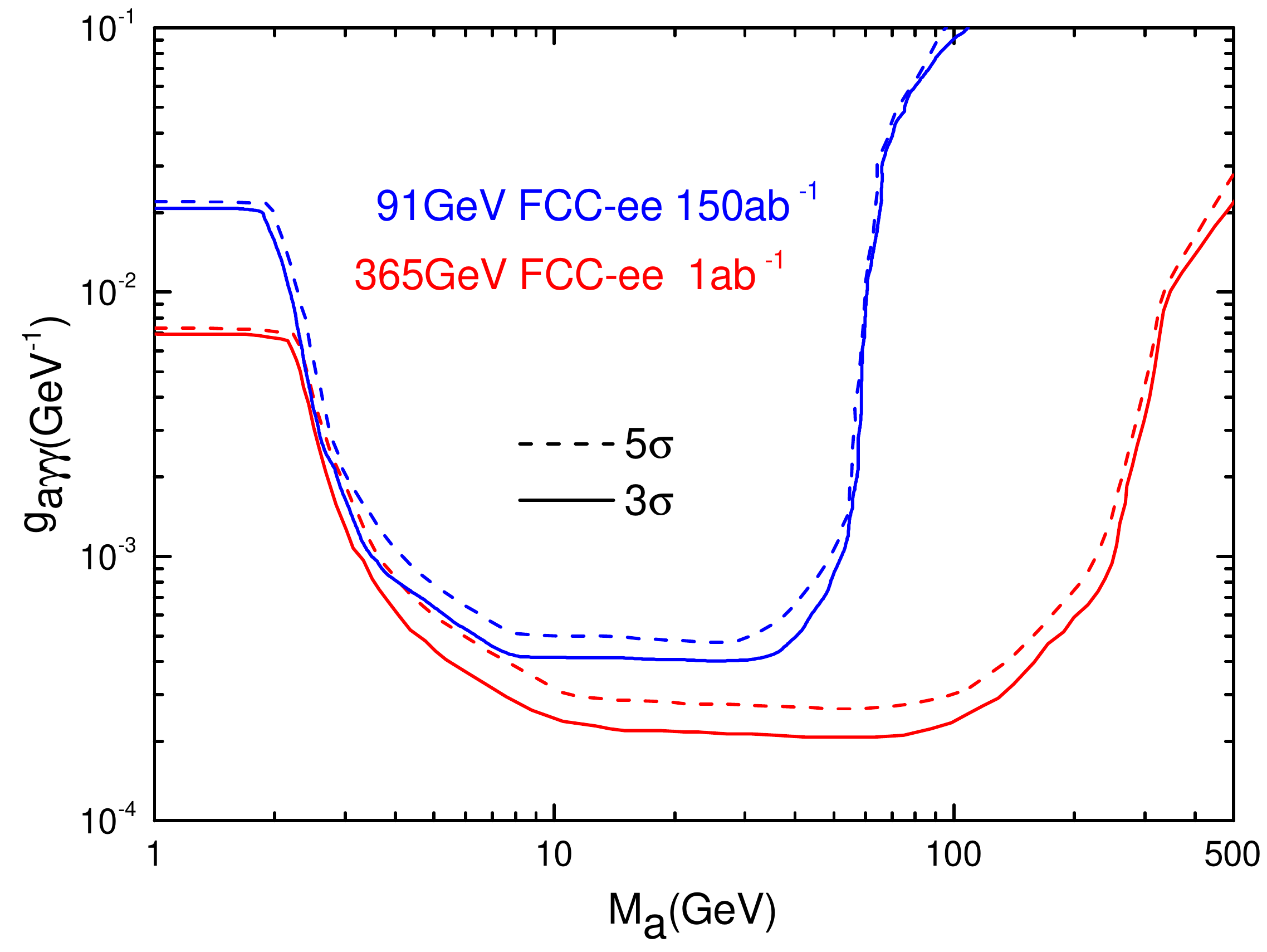}
		\caption{The 3$\sigma$ and 5$\sigma$ curves in the $M_{a}-g_{a\gamma\gamma}$ plane for $e^+e^-\rightarrow \gamma \gamma e^+e^-$ induced by $~~~~~~~~~$ALP at 365 GeV and 91 GeV FCC-ee with the designed luminosities.$~~~~~~$ }
		\label{FCC_2sigma}
	\end{center}
\end{figure}
In Fig.~\ref{FCC_2sigma}, we plot the 3$\sigma$ and 5$\sigma$ curves in the plane of $M_{a}-g_{a\gamma\gamma}$ for the FCC-ee with $\sqrt{s} $ = 365 GeV, $\mathscr{L}$ = 1.0 ab$^{-1}$ and $\sqrt{s} $ = 91 GeV, $\mathscr{L}$ = 150 ab$^{-1}$, respectively.
From Fig.~\ref{FCC_2sigma}, we can obtain that the sensitivity bounds as   $2.0{\times} 10^{-4}$ GeV${^{-1}} < g_{a\gamma\gamma} < 8 {\times} 10^{-3}$ GeV${^{-1}}$ ($2.8{\times} 10^{-4}$ GeV${^{-1}} < g_{a\gamma\gamma} < 1{\times} 10^{-2}$ GeV${^{-1}}$) in the ALP mass interval 1 GeV $\sim$ 340 GeV with $\sqrt{s}$ = 365 GeV   and $4.9{\times} 10^{-4}$ GeV${^{-1}} < g_{a\gamma\gamma} < 3.0{\times} 10^{-2}$ GeV${^{-1}}$ ($5{\times} 10^{-4}$ GeV${^{-1}} < g_{a\gamma\gamma} < 3.1{\times} 10^{-2}$ GeV${^{-1}}$) in the ALP mass interval 1 GeV $\sim$ 70 GeV  with $\sqrt{s} $ = 91 GeV  at 3$\sigma$ (5$\sigma$) levels, respectively. It is obvious that, comparing with 91 GeV FCC-ee, 365 GeV FCC-ee is more sensitive to ALP in the same mass range. In Fig.~\ref{FCC-LHC-CLIC}, our results for the sensitivities of the FCC-ee with $\sqrt{s}$ = 365 GeV and $\sqrt{s}$ = 91 GeV at 95$\%$ C.L. and other current exclusion regions for the coupling of ALP with photons are mapped into the $M_{a}-f^{-1}$ plane, where the relationship $\frac{g_{a\gamma \gamma}} 4 =f^{-1}$ is used.
 The current exclusion regions on the ALP-photon coupling from the results of searches for ALPs via the process $e^{+}e^{-}\to3\gamma$ at the LEP (light blue and dark blue) \cite{Acciarri:1994gb, Anashkin:1999yse, Mimasu:2014nea}, the same final states was studied at CDF (magenta) \cite{CDF} and LHC (peach) \cite{3gamma}. The green and light-shaded green regions respectively depict the results from searches for ALP via the LBL scattering at the LHC, ultraperipheral heavy-ion (Pb-Pb) collisions \cite{1607.06083, 1702.01625} and  the CLIC \cite{Inan:2020kif}. The exclusion regions also include the results from the central exclusive diphoton production at the LHC (light-shaded grey) \cite{Baldenegro:2018hng}, and the exotic decays $h \to Za$, $h \to aa$ and $Z \to \gamma a$ with $a$ decaying into a pair of charged lepton or two photons (lavender) \cite{ Bauer:2017ris}.
 \begin{figure}[!h]
 	\begin{center}
 		\includegraphics [scale=0.4] {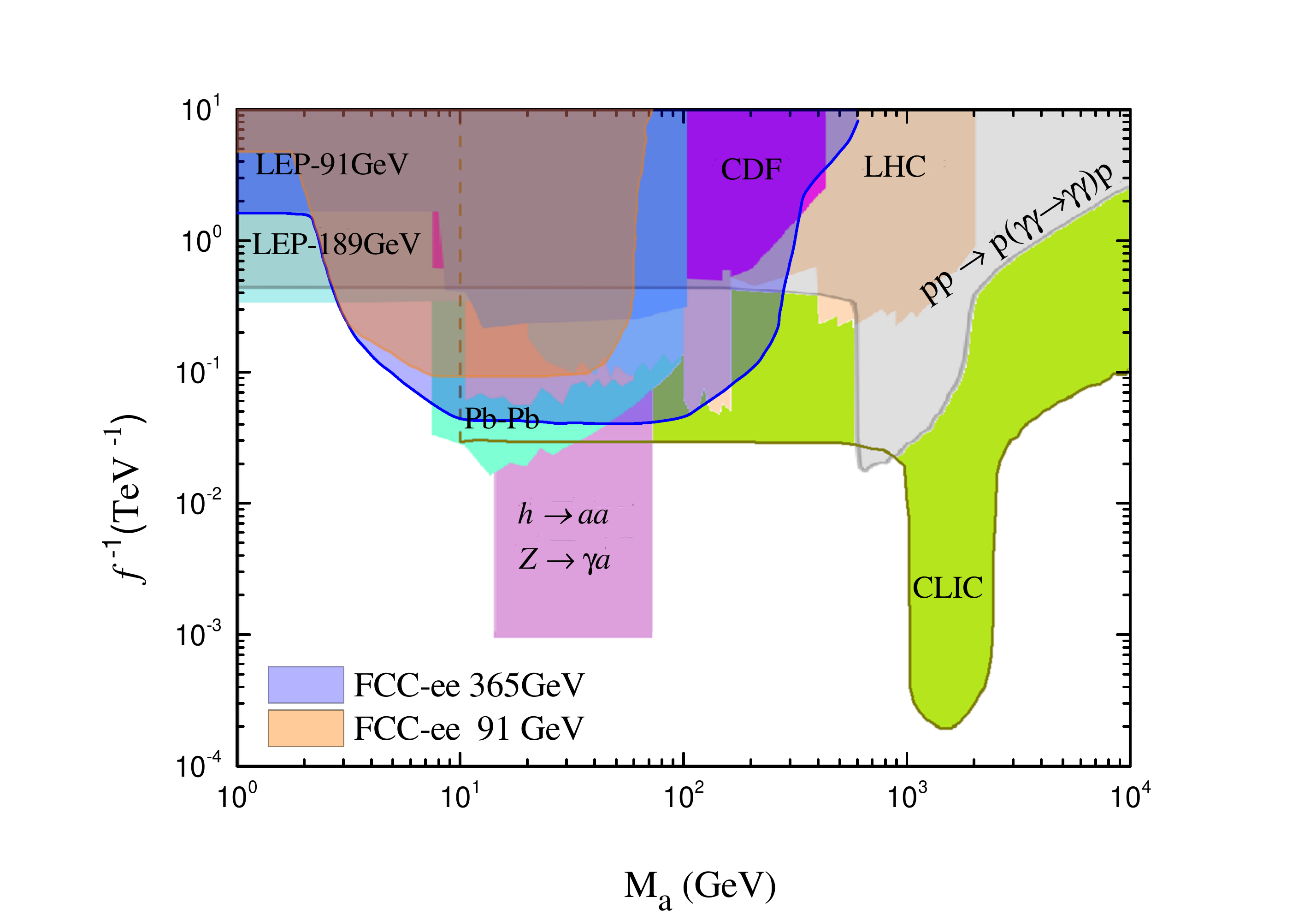}
 		\caption{The 95$ \% $ C.L. exclusion regions on the ALP couplings $g_{a\gamma\gamma}$ as function of $M_{a}$ from $~~~~$the process $e^+e^-\rightarrow \gamma \gamma e^+e^-$  at FCC-ee and other current exclusion regions.
 		}
 		\label{FCC-LHC-CLIC}
 	\end{center}
 \end{figure}
From Fig.~\ref{FCC-LHC-CLIC} we can see that the detectable mass ranges of the LHC~\cite{Baldenegro:2018hng} and CLIC~\cite{Inan:2020aal,Inan:2020kif} are much larger than those of the FCC-ee. This is because the higher collision energy can improve the detectability of new physics. For 8 GeV $\leq M_{a}\leq$ 300 GeV, our FCC-ee bounds on the ALP coupling with photons are stronger than those given by the LBL scattering at the LHC,  while are weaker than those of the polarized LBL scattering at the CLIC. Certainly, the ALP parameter space for this sector has been excluded by other experiments. However, for 2 GeV $\leq M_{a}\leq$ 8 GeV, the sensitivity bounds of 365 GeV FCC-ee with $\mathscr{L}$ = 1.0 ab$^{-1}$ are in the range of 0.06 TeV$^{-1} \sim$ 0.4 TeV$^{-1}$, which are stronger than those given by the LHC and CLIC. Thus, the FCC-ee might be more sensitive to the ALPs with mass 2 GeV $ \sim$ 8 GeV than the LHC and CLIC.

\section{The possibility of detecting ALPs  at CEPC}\label{(B)}
\begin{figure}[!ht]
	\centering
	\begin{subfigure}{0.19\linewidth}
		\includegraphics[width=\linewidth]{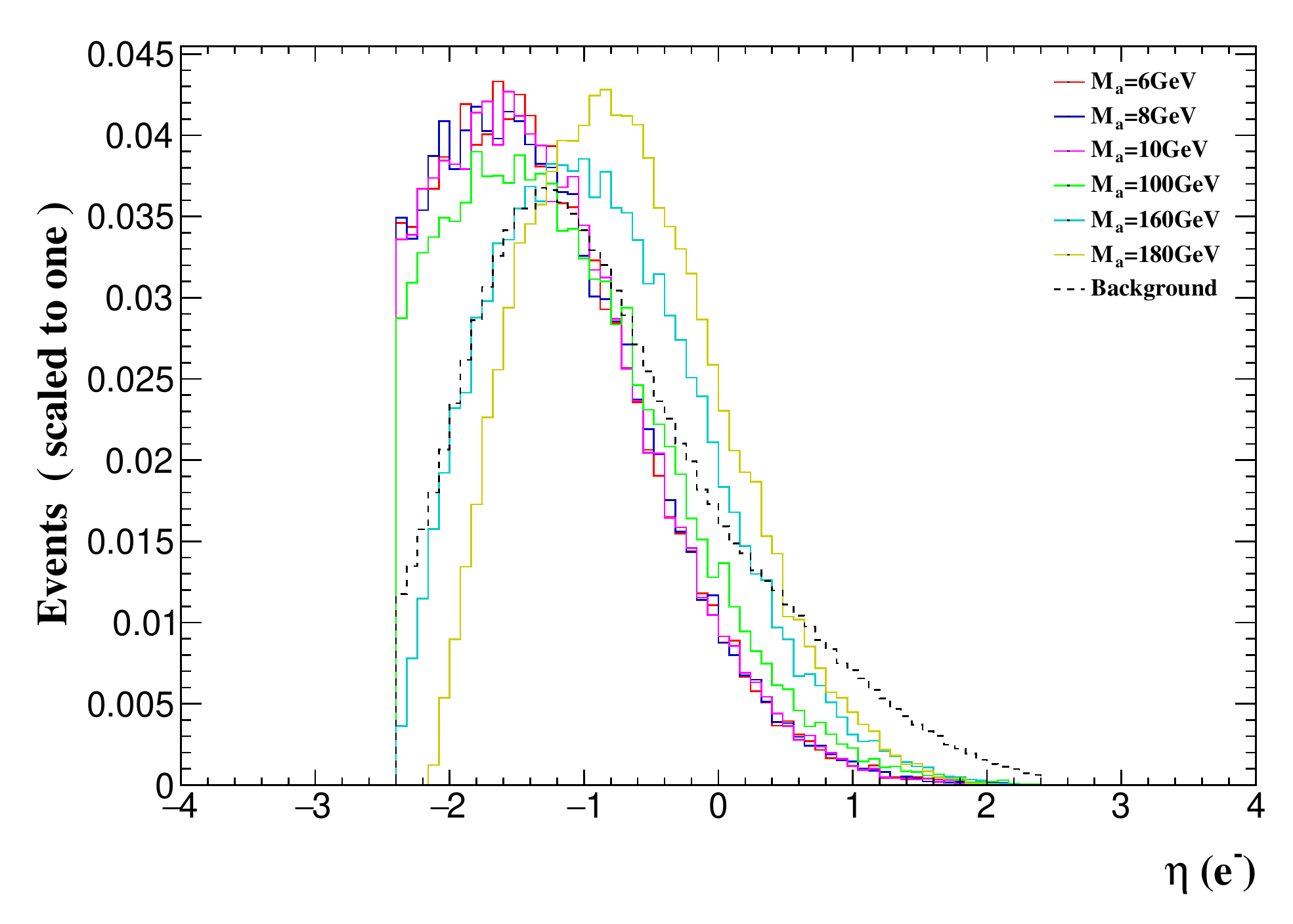}
		\caption{}\label{c}
	\end{subfigure}
	\begin{subfigure}{0.19\linewidth}
	\includegraphics[width=\linewidth]{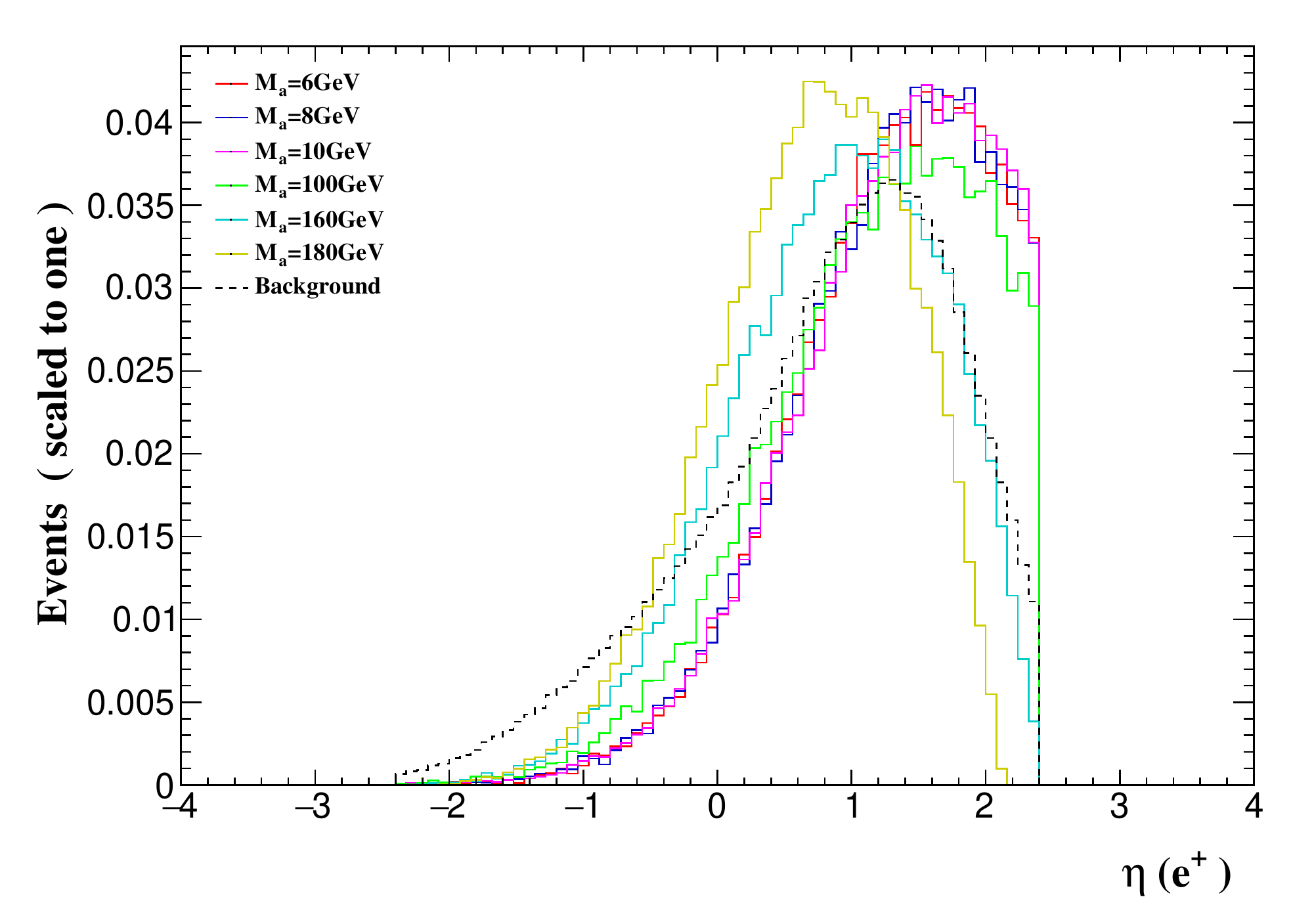}
	\caption{}\label{c}
	\end{subfigure}
	\begin{subfigure}{0.19\linewidth}
		\includegraphics[width=\linewidth]{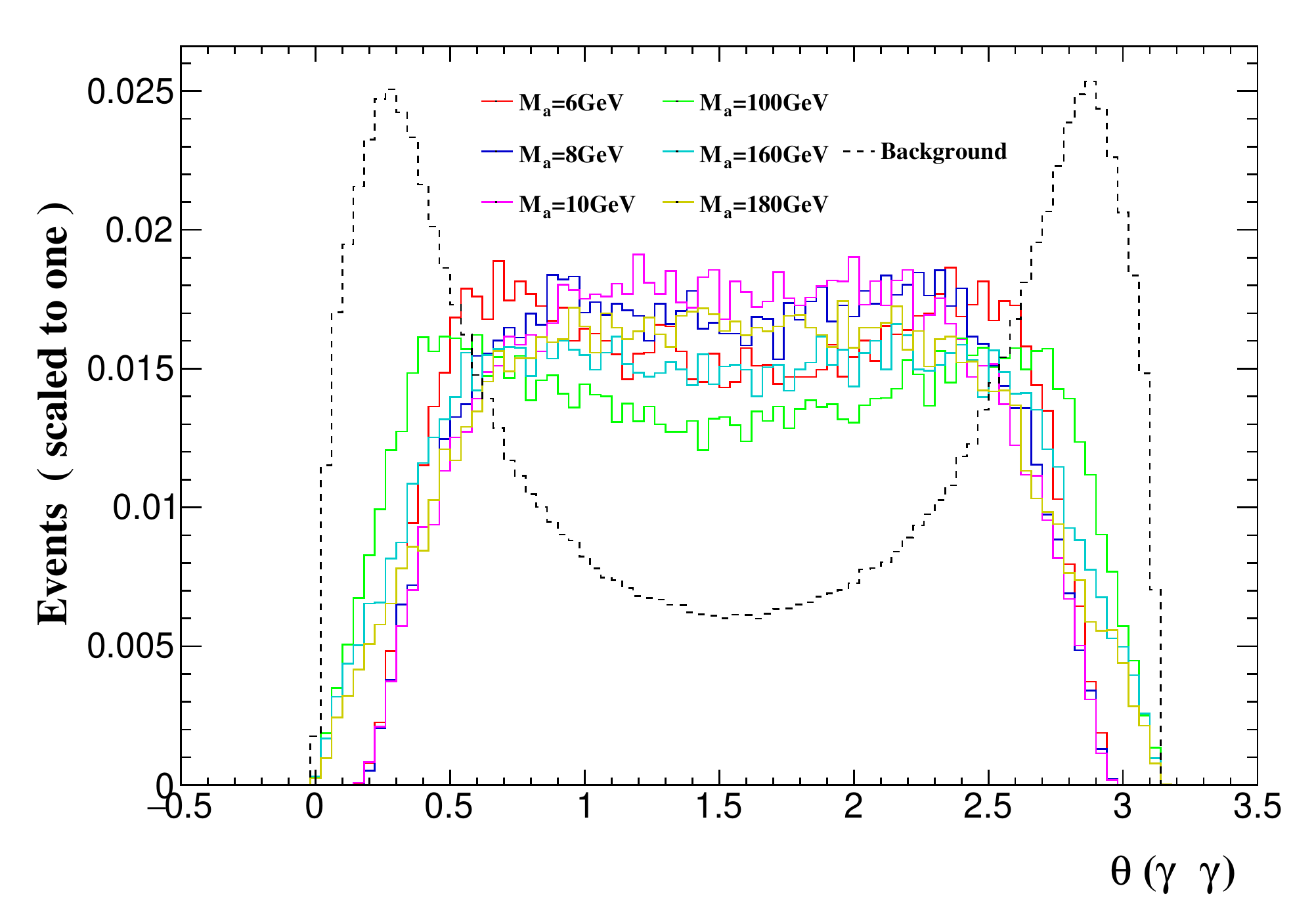}
		\caption{ }\label{a}
	\end{subfigure}
	\begin{subfigure}{0.19\linewidth}
	\includegraphics[width=\linewidth]{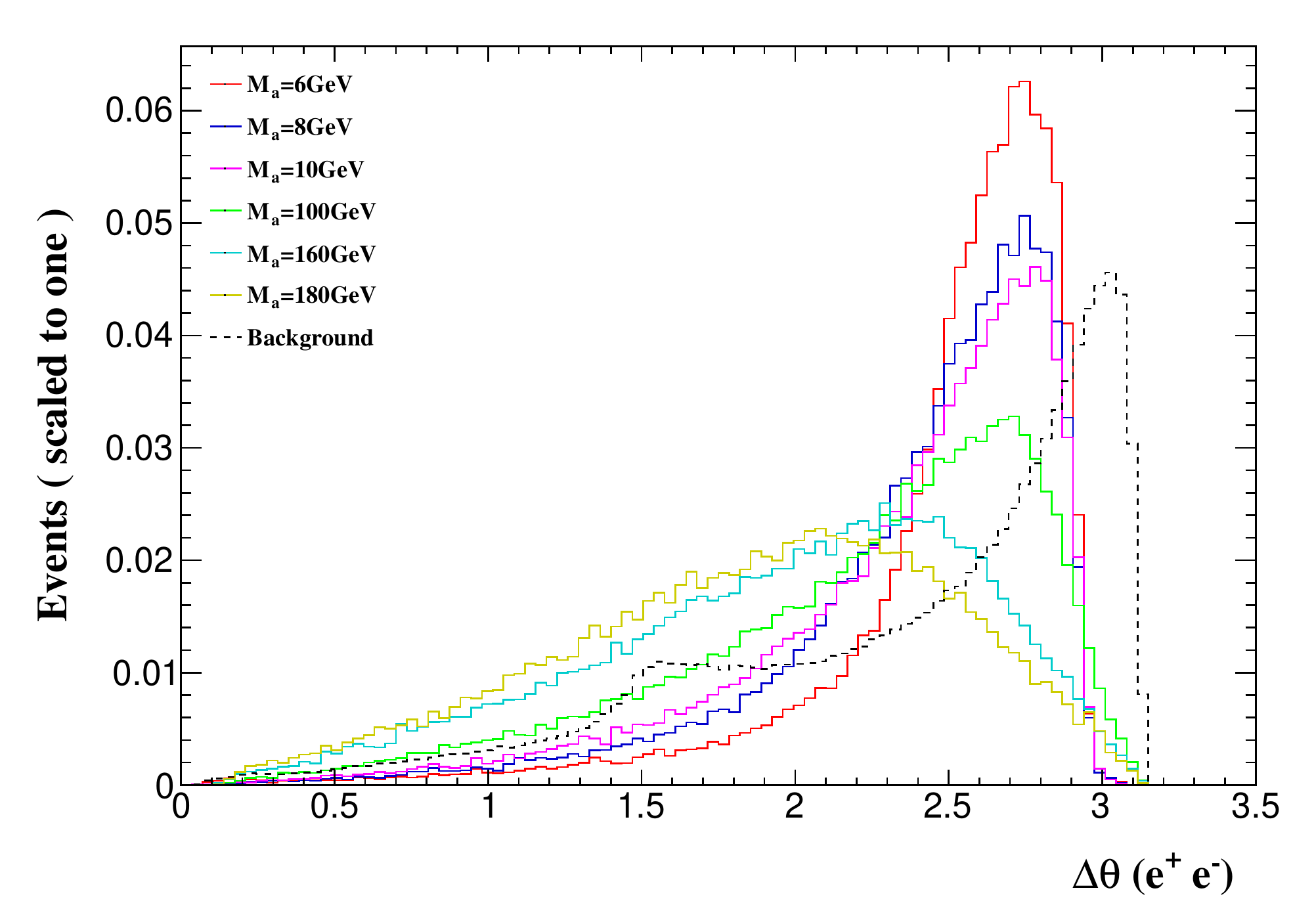}
	\caption{}\label{b}
	\end{subfigure}
	\begin{subfigure}{0.19\linewidth}
	\includegraphics[width=\linewidth]{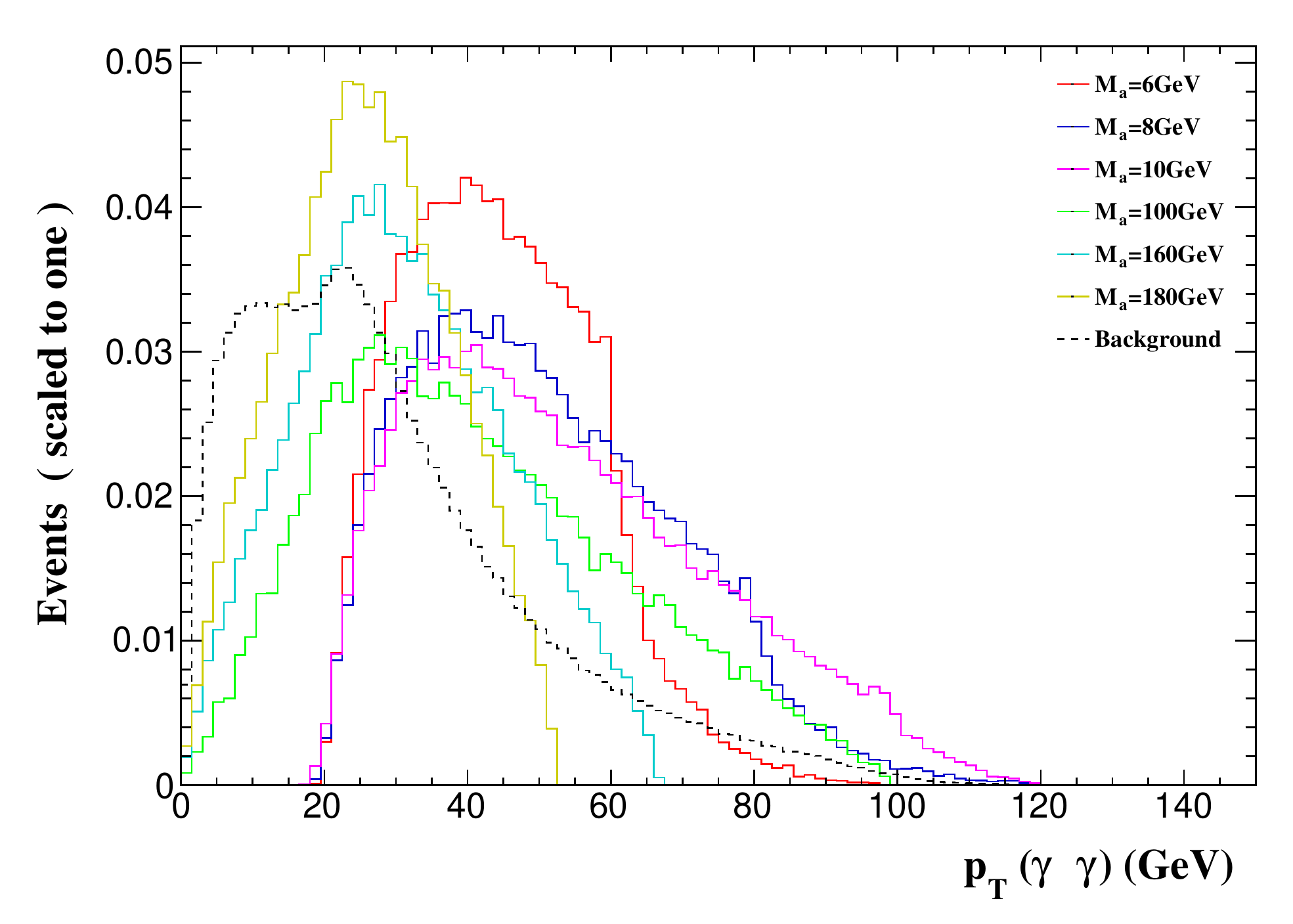}
	\caption{}\label{c}
	\end{subfigure}

\begin{subfigure}{0.19\linewidth}
	\includegraphics[width=\linewidth]{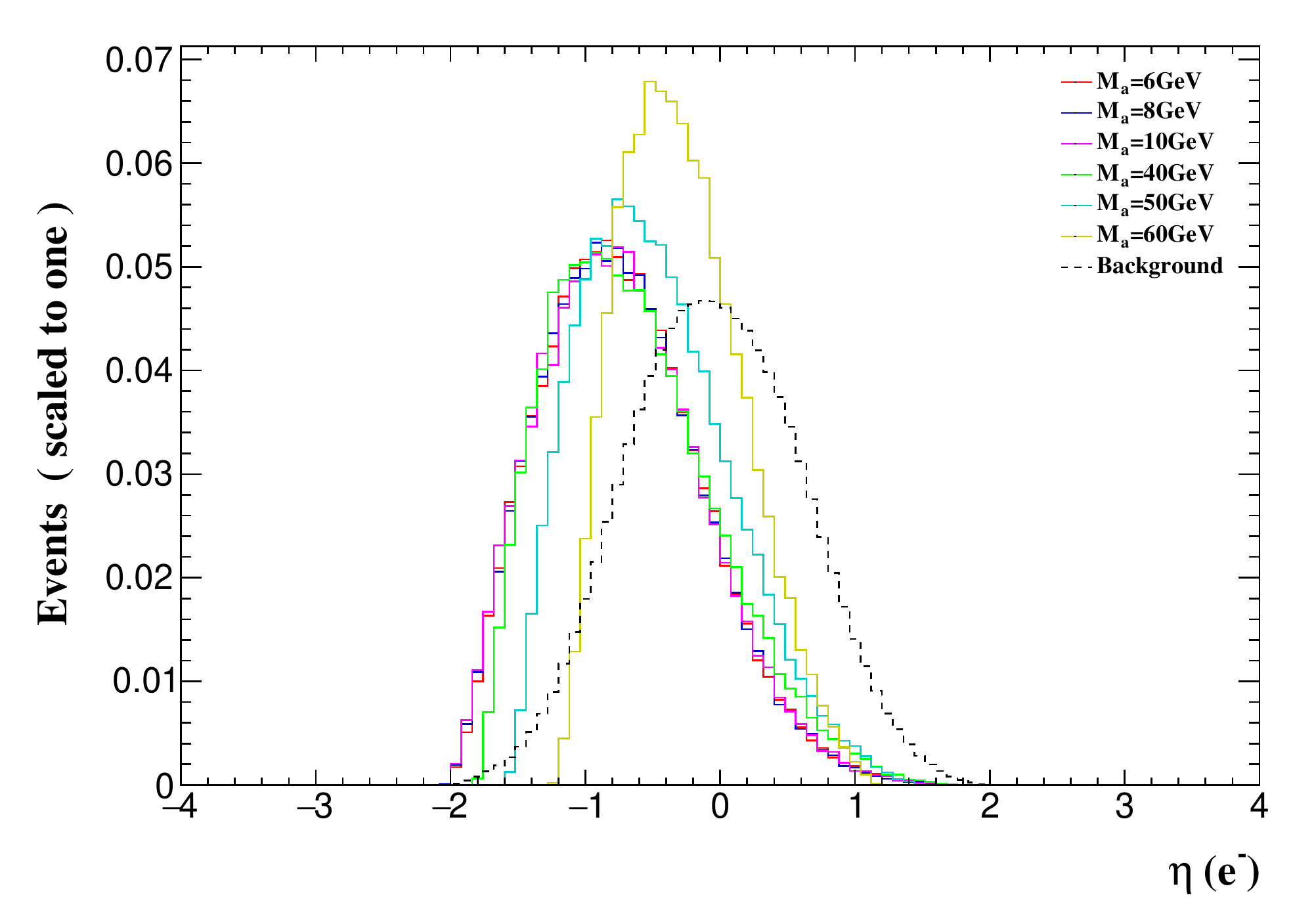}
	\caption{}\label{a.1}
\end{subfigure}
\begin{subfigure}{0.19\linewidth}
	\includegraphics[width=\linewidth]{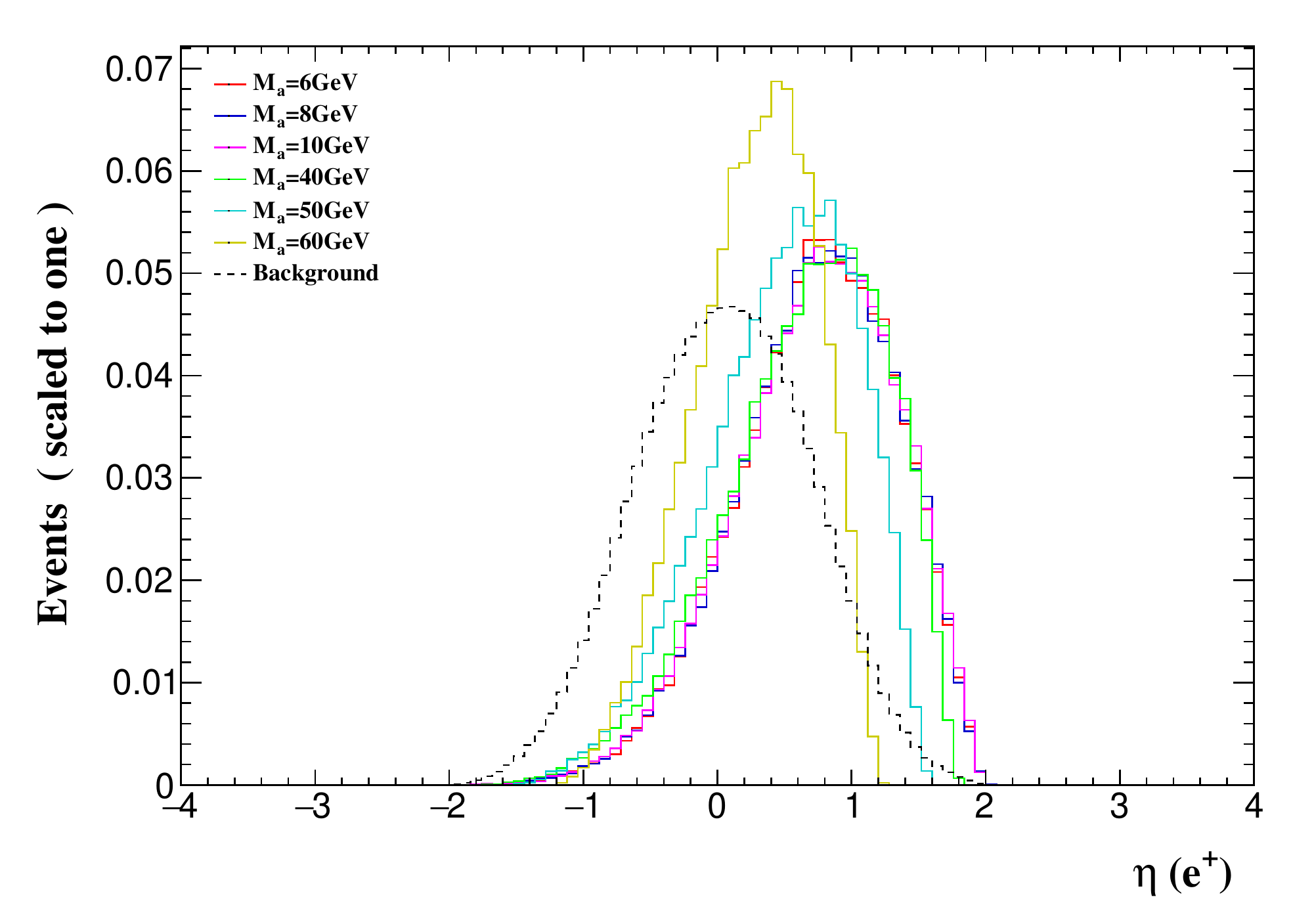}
	\caption{}\label{a.1}
\end{subfigure}
\begin{subfigure}{0.19\linewidth}
	\includegraphics[width=\linewidth]{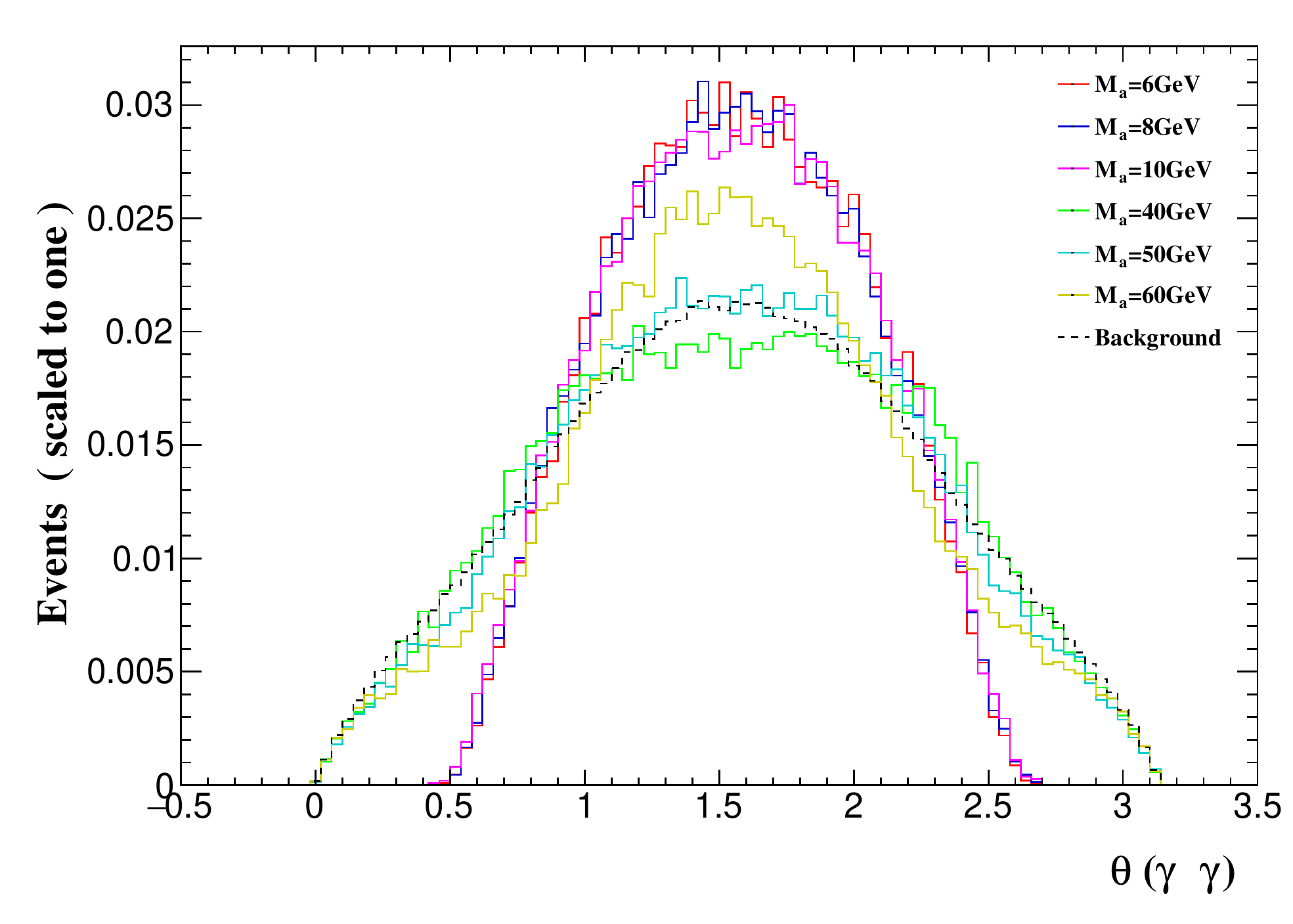}
	\caption{}\label{a.1}
\end{subfigure}
\begin{subfigure}{0.19\linewidth}
	\includegraphics[width=\linewidth]{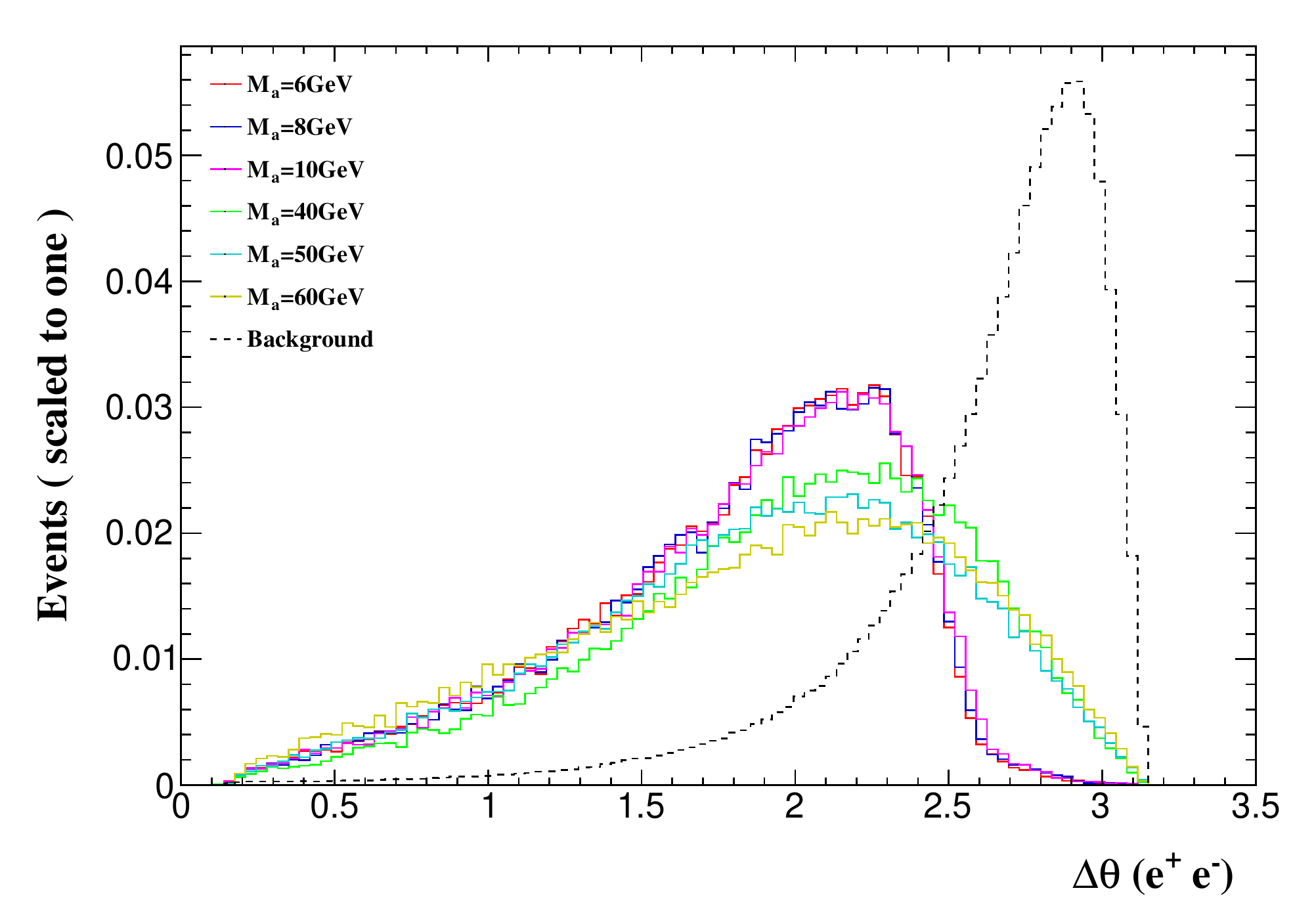}
	\caption{}\label{b.1}
\end{subfigure}
\begin{subfigure}{0.19\linewidth}
	\includegraphics[width=\linewidth]{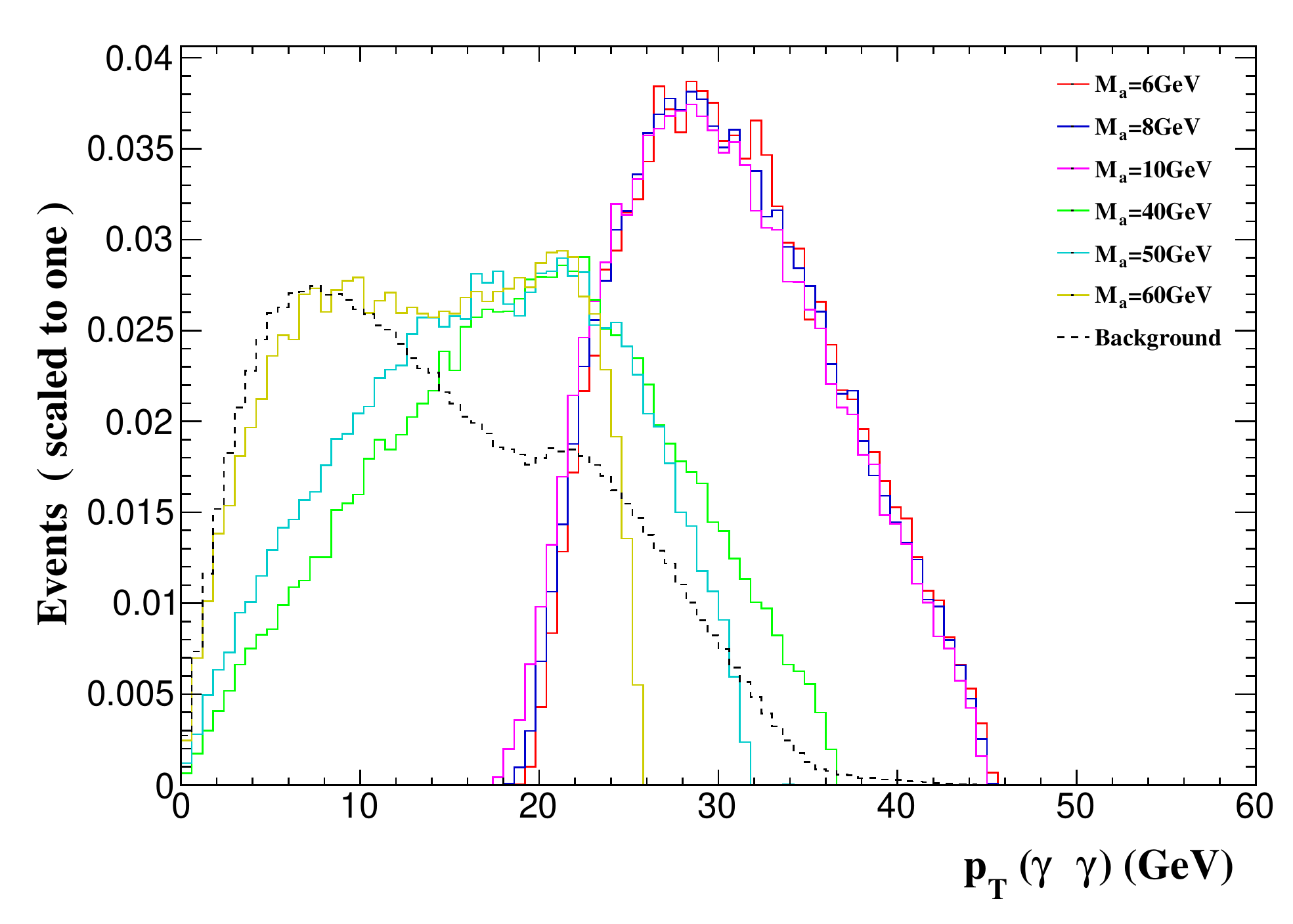}
	\caption{}\label{c.1}
\end{subfigure}
	\caption{Normalized distributions of $\eta {(e^{\pm})}$,~$\theta(\gamma \gamma),~\Delta\theta(e^{+} e^{-}),~p_{T} ({\gamma\gamma})$  for the signal of $~~~~~~~~~~~~~~~$selected ALP masses and the background at 240 GeV$~$ (a, b, c, d, e) and 91  $~~~ ~~$GeV (f, g, h, i, j) CEPC with the designed luminosities.$~~~~~~~~~~~~~~~~$ }\label{fig4}
\end{figure}

In this section, we use the similar method as in Sec.~\ref{(A)} to investigate the possibility of detecting ALPs at the CEPC with $\sqrt{s}$ = 240 GeV, $\mathscr{L}$ = 5.0 ab$^{-1}$  and  $\sqrt{s}$ = 91 GeV, $\mathscr{L}$ = 16 ab$^{-1}$  \cite{CEPCStudyGroup:2018rmc}.
 We perform the set of basic cuts (in Table \ref{basic cuts}) on the signal and background, and show the normalized distributions of   $\eta {(e^{\pm})}$, $\theta(\gamma \gamma)$, $\Delta\theta(e^{+} e^{-})$ and $p_{T} ({\gamma\gamma})$ of the signal and background in Fig.~\ref{fig4}.

We can see that the normalized distribution of $\theta(\gamma \gamma)$ and $\Delta\theta(e^{+} e^{-})$ in Fig.~\ref{fig4} are similar with those in Fig.~\ref{fig3}. Therefore, for $ \theta(\gamma \gamma)$ and $\Delta\theta(e^{+} e^{-})$, we choose the same cuts as in Sec.~\ref{(A)}.
The $ p_{T} ({\gamma\gamma}) $ distribution shows larger shift to the left compared to Fig.~\ref{fig3} (e), so we do not ignore the impact on cut efficiency and impose the new cut on $p_{T} ({\gamma\gamma}) $.
According to the above analysis, we choose the specific cuts as follows:
$$
\begin{array}{l}\scriptsize   %\footnotesize
	\centering{
		\newcolumntype{C}[1]{>{\centering\let\newline\\\arraybackslash\hspace{0pt}}m{#1}}
		\begin{tabular}{C{8cm}C{3cm}C{3cm} }
			Cuts    & $\sqrt{s}=$ 240 GeV &$\sqrt{s}=$ 91 GeV \\
			\multirow{2}{*}{\text { Cut-}1: Electron and positron pseudo-rapidity~~~~~~~~~~} &$0.4<\eta({e^{+})}<2.4 $&$-0.3<\eta({e^{+})}<0.9$ \\ 	
		                                                       	&$-2.4<\eta({e^{-})}<-0.4$&$-0.9<\eta({e^{-})}<0.3$\\
			\text { Cut-}2: Angle between the ALP and the beam axis~~~~ & 0.7~$\textless~\theta(\gamma \gamma)~\textless$~2.4   &~ 0.7~$\textless~ \theta(\gamma \gamma)~\textless$~2.4\\
			\text { Cut-}3: Angular separation between electron-positron &~~$\Delta\theta(e^{+} e^{-})~\textless$~2.9  &~~~  $\Delta\theta(e^{+} e^{-})~\textless$~2.4 \\
			\text { Cut-}4: Transverse momentum of reconstructed ALP~~ &~ $p_{T}({\gamma\gamma})~\textgreater~45~\mathrm{GeV}$ &~~~$p_{T}({\gamma\gamma})~ \textgreater~20~\mathrm{GeV}$  \\	
	\end{tabular}}
\end{array}
$$

\begin{figure}[ht!]
	\begin{center}
		\includegraphics [scale=0.4 ] {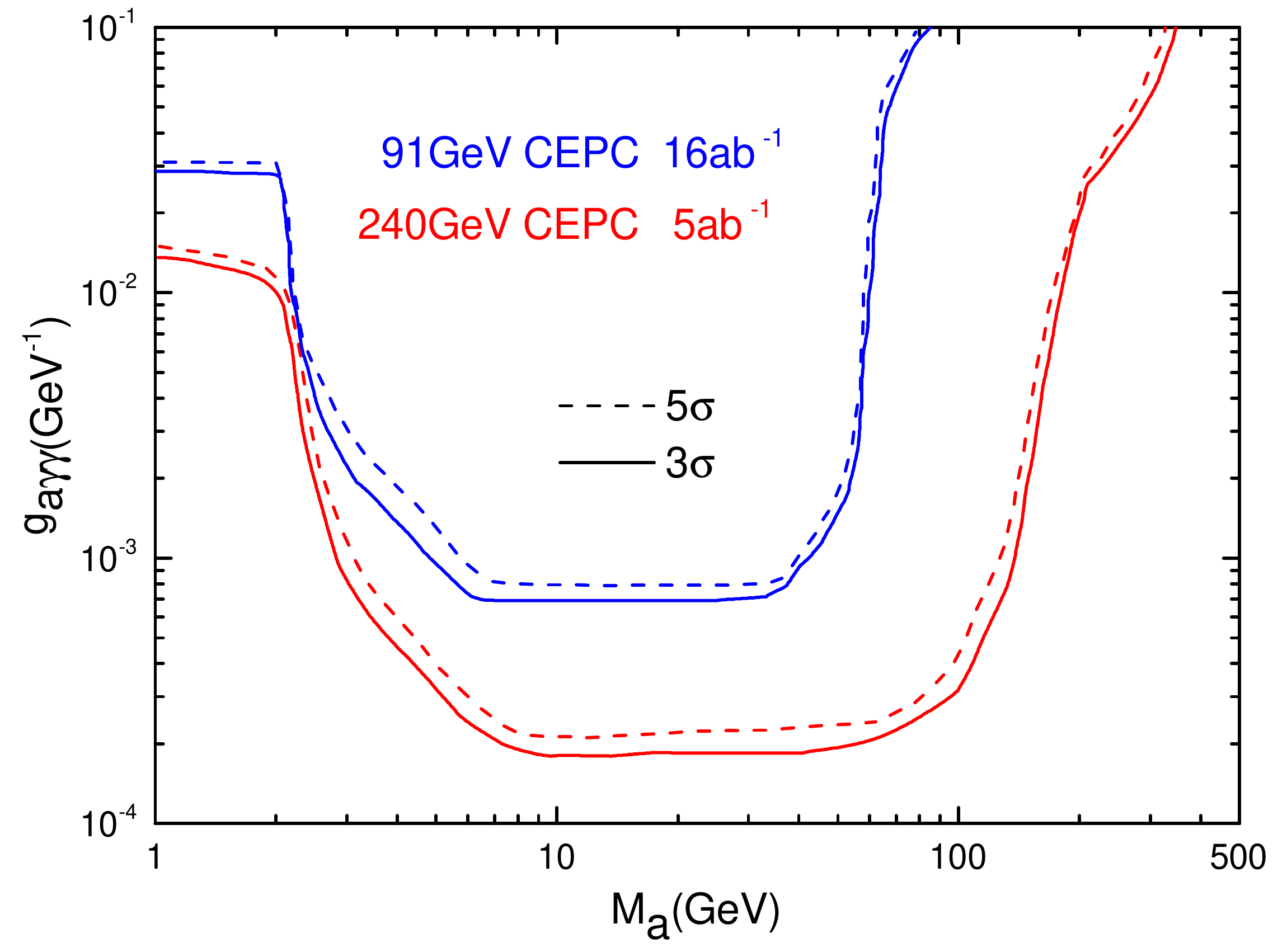}
		\caption{ The 3$\sigma$ and 5$\sigma$ curves in the $M_{a}$-$g_{a\gamma\gamma}$ plane for $e^+e^-\rightarrow \gamma \gamma e^+e^-$ induced by ALP at 240 GeV and 91 GeV CEPC with designed luminosities.$~~~~~~~~~~~~~~~~~$}\label{fig:K}
	\end{center}
\end{figure}

In Table~\ref{CEPC_table}, we summarize the numerical results of signal and background after imposing the above cuts at the CEPC.
From Table~\ref{CEPC_table}, we can see that the cross section of signal is reduced more than that of the  background after  imposing Cut-1 at $\sqrt{s} $ = 91 GeV. The signal reduces almost by a factor of 4-8 for low mass points (in the range of 2 GeV $< M_{a} <$ 8 GeV), while the background gets reduced by a factor of 1/2 (approximately). Therefore, we appropriately extend the limits of $\eta(e^{+})$ and $\eta(e^{-})$. Through calculation, Cut-1 is modified to $0.5<\eta(e^{+})<2 $ and $-2<\eta(e^{-})<-0.5$. After imposing the modified Cut-1, the cross section of signal reduces by about 35$\%$, and that of the background is reduced by 80$\%$. 
The SS can be improved 3-11 times with the optimized choice of cuts ($0.5<\eta(e^{+})<2 $ and $-2<\eta(e^{-})<-0.5$) compared to the existing choice of cuts for low mass ALPs at $\sqrt{s}=91$ GeV.
Even though the limits of $\eta(e^{+})$ and $\eta(e^{-})$ are extended, after imposing all cuts the SS at $\sqrt{s} $ = 91 GeV is still less than that at the CEPC with $\sqrt{s} $ = 240 GeV.
In Fig.~\ref{fig:K}, we plot the 3$\sigma$ and 5$\sigma$ curves in the plane of $M_{a}-g_{a\gamma\gamma}$ at the CEPC with $\sqrt{s}$ = 240 GeV, $\mathcal{L}$ = 5.0 ab$^{-1}$ and $\sqrt{s} $ = 91 GeV, $\mathscr{L}$ = 16 ab$^{-1}$, respectively. From Fig.~\ref{fig:K},  we can obtain that the sensitivity bounds as $1.8{\times} 10^{-4}$ GeV${^{-1}} < g_{a\gamma\gamma} < 2.1 {\times} 10^{-2}$ GeV${^{-1}}$ ($2.1{\times} 10^{-4}$ GeV${^{-1}} < g_{a\gamma\gamma} < 2.4{\times} 10^{-2}$ GeV${^{-1}}$) in the ALP mass interval 1 GeV $\sim$ 220 GeV  with $\sqrt{s} $ = 240 GeV and $7.0{\times} 10^{-4}$ GeV${^{-1}} < g_{a\gamma\gamma} < 6.0{\times} 10^{-2}$ GeV${^{-1}}$ ($8.2{\times} 10^{-4}$ GeV${^{-1}} < g_{a\gamma\gamma} < 6.1{\times} 10^{-2}$ GeV${^{-1}}$) in the ALP mass interval 1 GeV $\sim$ 70 GeV  with $\sqrt{s} $ = 91 GeV at 3$\sigma$ (5$\sigma$) levels, respectively. Similar to FCC-ee, 240 GeV CEPC is more sensitive to ALP than 91 GeV CEPC in the same mass range.

\begin{table}[h!]\tiny
	\centering	
	\newcolumntype{C}[1]{>{\centering\let\newline\\\arraybackslash\hspace{0pt}}m{#1}}
	\begin{tabular}{|C{1.2 cm}|C{1.8cm}|C{1.8cm}|C{1.8cm}|C{1.8 cm}|C{1.8 cm}|C{1.8cm}|C{2.0cm}| }
		\hline
		\multicolumn{8}{|c|}{ CEPC @ $\sqrt{s}=240$ (91) GeV }\\
		\hline
		\multirow{2}{*}{Cuts}    & \multicolumn{6}{c|}{Signal (fb) }&\multicolumn{1}{c|}{Background (fb) }   \\%\multicolumn{3}{c|}{SS}
		\cline{2-8}
		& $M_a$ = 6~GeV  & $M_a$ = 8~GeV  & $M_a$ = 10~GeV  & $M_a$ = 50~GeV  & $M_a$ = 100~GeV   & $M_a$ = 160~GeV &${\gamma \gamma e^{+} e^{-}}$  \\
		\hline
		Basic cuts & 3.4378(0.249) & 4.8088(0.4796) & 5.2928(0.5003)    & 5.9064(0.2432)     & 3.585  & 0.8021     & 67.0614(98.8986)  \\
		Cut 1      &2.9865(0.0316)&3.932(0.1267) &4.138(0.1417)    &4.5336(0.0977)     &2.4778  &0.4436     &33.7026(40.928) \\
		Cut 2      &2.1714(0.0309)&3.0176(0.1264)&3.2819(0.1411)    &3.1262(0.0904)   &1.6993  & 0.3145    &12.628(34.93)\\
		Cut 3      &2.1368(0.0226)&3.0383(0.1156)&3.2422(0.1297)   &3.0238(0.0717)    &1.6497  &0.3052     &9.042(8.396)\\
		Cut 4      &1.4(0.0226)   &2.2984(0.1156)&2.5065(0.1297)   &2.0519(0.0501)     &0.8747  &0.0392     &3.3614(6.1921)\\
		\hline	
	\end{tabular}
	\caption{After different cuts applied, the cross sections for the signal and background $~~~~~~~~~~~~~~~~~~$at 240 (91) GeV CEPC with $g_{a\gamma\gamma} = 10^{-3}$ GeV$^{-1}$ and the benchmark points $M_{a}$ = 6, 8, 10, 50, 100, 160 GeV (6, 8, 10, 50 GeV).$~~~~~~~~~~~~~~~$\label{CEPC_table}
}
\end{table}
%%%%%%%%%%%%%%%%%%%%%%%%%%%%%%%%%%%%%%%%%%%%%%%%%%%%%%%%%%%%%%%%%%%%%%%%%%%%%%%%%%%%%%%%%%%%%%%%%%%%%%%
%%%%%%%%%%%%%%%%%%%%%%%%%%%%%%%%%%%%%%%%%%%%%%%%%%%%%%%%%%%%%%%%%%%%%%%%%%%%%%%%%%%%%%%%%%%%%%%%%%%%%%%

\begin{figure}[!h]
	\begin{center}
		\includegraphics [scale=0.4] {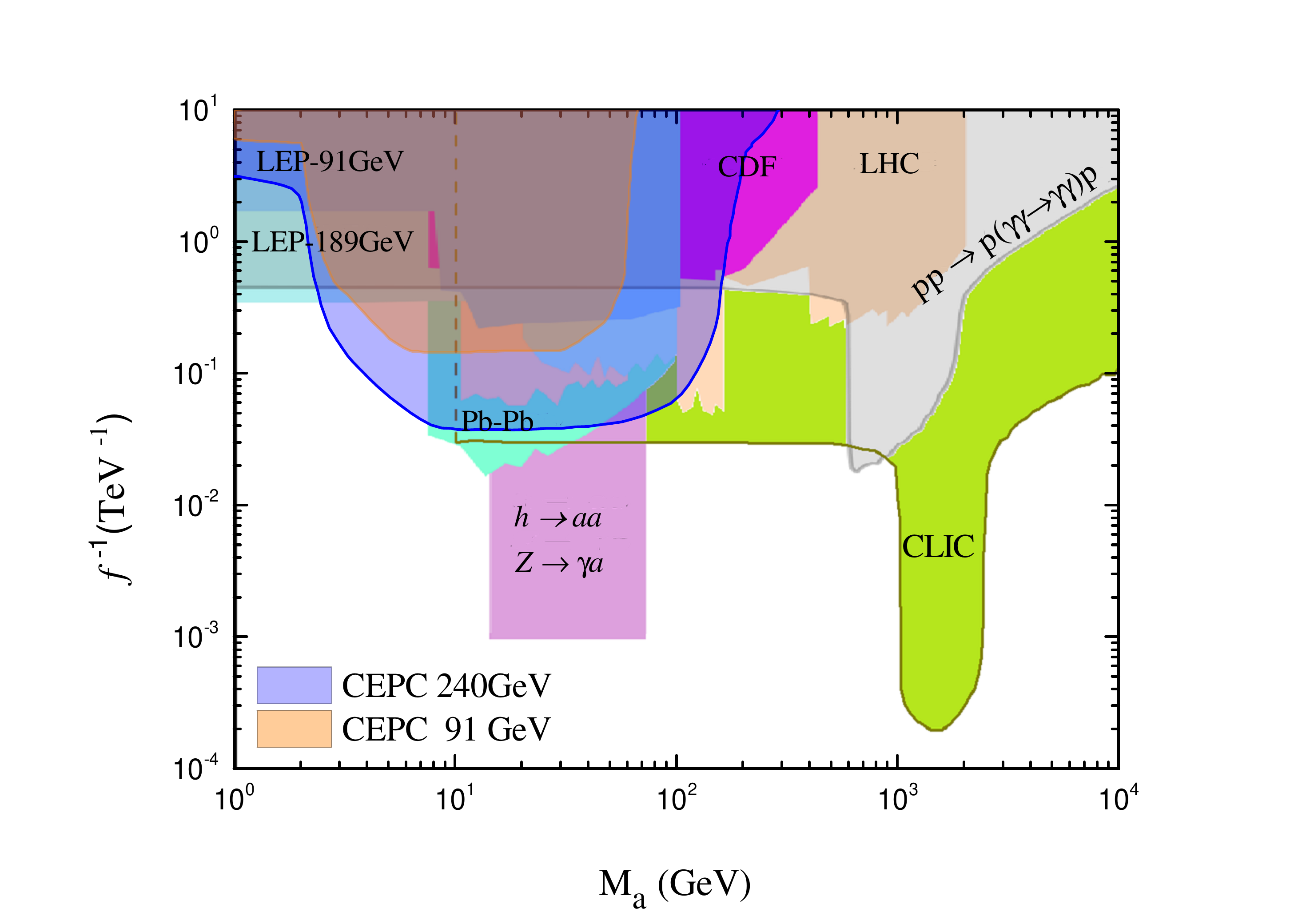}
		\caption{The 95$ \% $ C.L. exclusion regions on the ALP couplings $g_{a\gamma\gamma}$ as function of $M_{a}$ from $~~~~~~~$the process $e^+e^-\rightarrow \gamma \gamma e^+e^-$ at the CEPC and other current exclusion regions. 
		}
		\label{CEPC-LHC-CLIC}
	\end{center}
\end{figure}
%%%%%%%%%%%%%%%%%%%%%%%%%%%%%%%%%%%%%%%%%%%

In Fig.~\ref{CEPC-LHC-CLIC}, the sensitivities of the CEPC with $\sqrt{s}$ = 240 GeV and 91 GeV  at 95$ \% $ C.L.  and other current exclusion regions for the coupling of ALP with photons are mapped into the $M_{a}-f^{-1}$ plane.
Compared with the results of the LHC~\cite{Baldenegro:2018hng} and CLIC~\cite{Inan:2020aal,Inan:2020kif}, the CEPC has better potential for detecting the ALP coupling with photons at GeV scale.
For 2 GeV $\leq M_{a}\leq$ 8 GeV, the sensitivity bounds of the 240 GeV CEPC with $\mathscr{L}$ = 5.0 ab$^{-1}$ are in the range of 0.05 TeV$^{-1} \sim $ 0.4 TeV$^{-1}$, which are even stronger than those of the FCC-ee slightly.
Thus, the CEPC might be the most sensitive to the ALPs with mass 2 GeV $\sim$ 8 GeV than other colliders.

\section{Conclusions and discussions}\label{conclusions} 
As pseudo-Goldstone bosons, ALPs are one of the most potential particles to actually exist. There will be great expectations for discovering them in future electron-positron colliders.
In this paper, we have investigated the observability of the ALP diphoton signal through the process $e^+e^-\rightarrow \gamma \gamma e^+e^-$  at the FCC-ee and CEPC.
Our results show that the detectable mass ranges of the FCC-ee and CEPC are much smaller than those of the LHC and CLIC.
For 8 GeV $\leq M_{a}\leq$ 300 GeV, our FCC-ee and CEPC bounds on the ALP coupling with photons are stronger than those given by the LBL scattering at the LHC,  while are weaker than those of the polarized LBL scattering at the CLIC. However, for 2 GeV $\leq M_{a}\leq$ 8 GeV, the sensitivity bounds of 365 GeV FCC-ee with $\mathscr{L}$ = 1.0 ab$^{-1}$ and 240 GeV CEPC with $\mathscr{L}$ = 5.0 ab$^{-1}$ are respectively in the ranges of 0.06 TeV$^{-1} \sim $ 0.4 TeV$^{-1}$ and  0.05 TeV$^{-1} \sim $ 0.4 TeV$^{-1}$, which are stronger than those given by the LHC and CLIC. Thus, both FCC-ee and CEPC might be more sensitive to the ALPs with mass 2 GeV $ \sim$ 8 GeV than the LHC and CLIC.

A light ALP can arise as a long-lived particle and might lead to a pair of photons from a displaced vertex \cite{Mimasu:2014nea}. The LHC has  performed a search for displaced vertices from long-lived particle in the inner detector and muon system \cite{Agrawal:2021dbo}. The experimental limits can give constraints for two $M_a$ ranges: $2m_\mu<M_a<m_B-m_K$ and $2m_\mu<M_a<m_K-m_\pi$ \cite{Gavela:2019wzg}.
This range covers part of the parameter space we are discussing, in which the ALP might yields displaced vertices.
In our work, the decay width of ALP is about $10^{-8}\sim10^{-10}$ GeV for 2 GeV $\leq M_a \leq 8$ GeV and $g_{a\gamma\gamma}\sim 0.1$ TeV$^{-1}$. The proper lifetime $\tau_a$ is less than $10^{-3}$ ps. Considering that ALP would decay promptly for $\tau_a < 1$ ps \cite{Gavela:2019wzg}, The LBL scattering induced by ALP is not affected by displaced vertices.

In conclusion, our results represent a viable way to detect the ALPs with low masses though the LBL scattering. As a complementary to study of ALPs, LBL scattering processes at the FCC-ee and CEPC provide a new possibility to search ALPs in future electron-positron collider experiments.

\textbf{Note added.}
When finishing this work, we find that Ref.~\cite{Goncalves:2021pdc} derives the expected sensitivity to the ALP with mass 2 GeV $ \sim$ 5 GeV in ultraperipheral $Pb-p$ and $Pb - Pb$ collisions at the LHC.

\section*{Acknowledgement}
This work was supported in part by the National Natural Science Foundation of China under Grants No.11905093, 11875157 and 12047570; and by the Doctoral Start-up Foundation of Liaoning Province No.2019-BS-154.

\end{document}